\documentclass{article}
\usepackage{arxiv}
\usepackage{fontenc}
\usepackage[english]{babel}
\usepackage{caption}
\usepackage{lscape} 
\usepackage{amsmath}
\usepackage{geometry}
\usepackage{amsthm}
\usepackage{adjustbox}
\usepackage{amsfonts}
\usepackage{mathtools}
\usepackage{verbatim}
\usepackage{algpseudocode,algorithm,algorithmicx}
\usepackage{mathtools}
\usepackage{comment}
\usepackage{verbatim, times, booktabs, dcolumn, setspace, fullpage}

\usepackage{graphicx}
\usepackage{wrapfig}
\usepackage{caption}
\usepackage{mathtools}
\usepackage{float}

\usepackage[english,noautotitles-r]{SASnRdisplay} 
\usepackage{url}
\usepackage{subfigure}
\lstdefinestyle{r-output}{
style = r-style,
style = r-output-user,
}

\usepackage{url}            
\usepackage{booktabs}       
\usepackage{amsfonts}       
\usepackage{nicefrac}       
\usepackage{microtype}      
\usepackage{graphicx}
\usepackage{doi}

\title{A Laplace transform-based test for the equality of positive semidefinite matrix distributions

}

\author{ \href{https://orcid.org/0000-0002-1964-7539}{\includegraphics[scale=0.06]{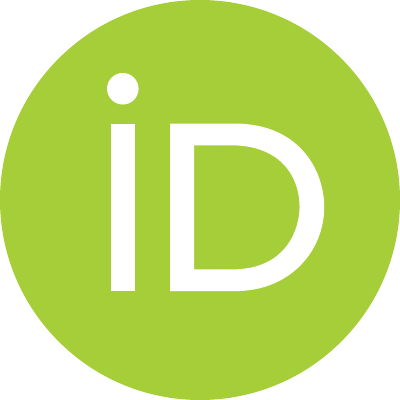}\hspace{1mm} Žikica Lukić} \\
	PhD student at the Faculty of Mathematics\\
	University of Belgrade\\
	Belgrade, 11000, Serbia \\
	\texttt{zikicamaster@gmail.com} \\
}

\begin{document}

\maketitle

\begin{abstract}
In this paper, we present a novel test for determining equality in distribution of matrix distributions. Our approach is based on the { integral} squared difference of the empirical Laplace transforms with respect to the noncentral Wishart measure. We conduct an extensive power study to assess the performance of the test and determine the optimal choice of parameters. Furthermore, we demonstrate the applicability of the test on financial and non-life insurance data, illustrating its effectiveness in practical scenarios.
\end{abstract}



{\section{Introduction}\label{sec::intro}}

The field of matrix statistics has been a rapidly developing field due to advances in computational methods and numerous real-world applications. 

Matrix-variate distributions { are an extension of the concept of the multivariate distribution to matrices. These distributions allow for the modelling of covariances and dependencies among multiple variables simultaneously. Matrix distributions} have been applied  { in various fields. In finance, they are used to model volatility on the financial markets \cite{gourieroux2010derivative},  detect changes in the cryptocurrency markets \cite{nas1}, and model the portfolio returns \cite{alfelt2023}. In medical research, the applications are mainly devoted to diffusion tensor imaging (DTI) \cite{dryden2009, jian2007multi, jian2007novel}. DTI is a novel magnetic resonance imaging method that has attracted extensive interest,
especially due to its applications in the study of brain diseases \cite{hadjicostaThesis}. Matrix distributions are also applied in wireless communication systems \cite{siriteanu2016chi}.

The most prominent representatives of the symmetric positive definite matrix distributions in matrix statistics are the Wishart and the noncentral Wishart distributions. The Wishart distribution can be seen as a matrix extension of the $\chi^2$ distribution.  Some results related to statistical testing which utilize the Wishart distribution can be found in \cite{hadjicosta2020integral, nas1}.
The results concerning the testing which utilizes the centralized Wishart process can be found in \cite{alfelt2020testing}.

Analogously, the noncentral Wishart distribution is the matrix extension of the noncentral $\chi^2$ distribution. The distributional properties of the noncentral Wishart distribution have been extensively studied in the literature \cite{letac2018laplace, mayerhofer2013existence,   mayerhofer2019wishart}.}

It is worth mentioning that integral transform methods{, such as Hankel transform methods \cite{hadjicosta2020integral, nas1, nas2},} have recently gained popularity in matrix statistics. However, research in the matrix-variate case appears to be limited when compared to the univariate case.

In the one-dimensional case, several integral transforms have been utilized in the construction of statistical tests. Hankel transform goodness-of-fit (GOF) tests have been developed for the exponential distribution \cite{baringhaus2010empirical, baringhaus2013ks} and the gamma distribution \cite{hadjicosta2020Gamma}. Laplace transforms have been employed in the development of GOF tests for the inverse Gaussian distribution \cite{henze2002IG}, Rayleigh distribution \cite{meintanis2003tests},  exponential distribution \cite{cuparic2022new, henze2002tests, milovsevic2016new}, L\'evy distribution \cite{lukic2023characterization}, and gamma distribution \cite{henze2012goodness}. Additionally, a two-sample test of equality in distribution utilizing empirical Hankel transforms has been explored in \cite{baringhaus2015two}.

The methodologies employed in the univariate case can be extended to the matrix-variate scenario, as matrix Laplace transforms share many analogous properties with their one-dimensional counterparts. It is well known that the Laplace transform of a matrix random variable uniquely determines its distribution \cite{farrell1985multivariate}, and the formulas for Laplace transforms of many matrix-variate distributions are known in the literature (see \cite{gupta1999matrix, muirhead2009aspects}).

However, to the best of our knowledge, the matrix Laplace transforms have not been utilized in the construction of either GOF tests or tests of equality in distribution in the literature.

Therefore, to fill the existing gap, it is natural to follow the approach considered in \cite{nas1} and construct the test of equality in distribution as an integral squared difference of two empirical Laplace transforms with respect to the suitably chosen measure.

The main objective of this paper is to expand upon the research presented in \cite{nas1} by developing an integral two-sample test for testing equality between two arbitrary positive definite matrix-valued distributions, regardless of their orthogonal invariance. The test statistic will be constructed using the properties of the empirical Laplace transforms with respect to the noncentral Wishart measure. 

The paper is organized as follows: { Section \ref{sec::existing} provides an overview of existing tests for matrix data that employ integral transform methods.} In Section \ref{sec::teststat}, { the novel} test statistic is presented. Section \ref{sec::power} is dedicated to the power study, while Section \ref{sec::realdata} contains real data examples that demonstrate the application of the proposed methodology. { Appendix I contains the empirical 95-th percentiles of the distribution of the scaled test statistic.}

\medskip

\medskip

{ \section{Existing integral transform-type tests for matrix data}\label{sec::existing}}

{
In this section, we introduce the two existing tests for matrix data that employ integral transforms and detail the construction of the corresponding test statistics.
}

The (orthogonally invariant) Hankel transform has been utilized to construct a GOF test of Wishart distribution \cite{hadjicosta2020integral}. 
The authors have constructed a test statistic as an integrated square difference of the empirical orthogonal Hankel transform and the Hankel transform of the Wishart distribution. 

Denote with $T > 0$ the matrix is positive definite, $tr(X)$ denotes a trace of the matrix, and denote with $dW(T)$ a standard Wishart measure. Unless otherwise specified, we assume that the matrices are $d\times d$ real matrices.
Then the test { statistic} proposed by Hadjicosta and Richards can be expressed as:
\begin{equation*}
    T_n^2 = n\int\limits_{T>0} [\Tilde{\mathcal{H}}(T)-\exp(tr(-T/\alpha))]^2 dW(T).
\end{equation*}

Here, $\Tilde{\mathcal{H}}(T)$ represents the empirical orthogonal Hankel transform of order $\nu$, defined as:

\begin{equation*}
   \Tilde{\mathcal{H}}_{n, \nu} (T)=\Gamma_d(\nu+\frac{1}{2}(d+1))\frac{1}{n}\sum\limits_{j=1}^{n} A_\nu (T, X_j).
\end{equation*}

In the above expressions, $\Gamma_d(\cdot)$ denotes the multivariate gamma function, and $A_\nu (\cdot, \cdot)$ represents the Bessel function of two matrix arguments. { The rationale behind this construction is that the empirical orthogonal Hankel transform is a consistent estimator of the orthogonal Hankel transform.
The orthogonal Hankel transform uniquely characterizes the distribution. Therefore, if the data $X_1, X_2, \dots, X_n$ come from the Wishart distribution, the value of the square difference $[\Tilde{\mathcal{H}}(T)-\exp(tr(-T/\alpha))]^2$ is expected to be small, and thus the value of the test statistic is expected to be small. This approach is common in the construction of the integral transform type GOF statistical tests. For some recent developments in the univariate case, see e.g. \cite{bojanaEYSM, bojanaSimos}.

For the explicit form of the test statistic and further properties of the test,} please refer to \cite{hadjicosta2020integral}. 
{ This is the only known GOF test in the space of symmetric positive definite matrix distributions up to date, employing the integral transform method.

For further information on integration in the space of symmetric positive definite matrices, please refer to \cite{farrell1985multivariate, muirhead2009aspects}.}

Following the { construction} presented in \cite{hadjicosta2020integral}, a two-sample test of equivalence within the class of matrix distributions orthogonally invariant in distribution was developed in \cite{nas1}. The test statistic was constructed as the integrated squared difference between two orthogonally invariant Hankel transforms. { Given the samples $\mathbf{X}=X_1, X_2, \dots, X_{n_1}$ and $\mathbf{Y}=Y_1, Y_2, \dots, Y_{n_2}$},  the test statistic { can be} formulated as:
\begin{equation}\label{NasStari}
    I_{n_1, n_2,  \nu}=\int\limits_{T>0}\Big(\Tilde{\mathcal{H}}_{{n_1}, \nu} (T)-\Tilde{\mathcal{H}}_{{n_2}, \nu} (T)\Big)^2dW(T).
\end{equation}

Here, $\Tilde{\mathcal{H}}_{n_1, \nu} (T)$ and $\Tilde{\mathcal{H}}_{n_2, \nu} (T)$ represent the empirical orthogonal Hankel transforms of the { samples $\mathbf{X}$ and $\mathbf{Y}$}, respectively. 
{ The rationale behind this construction is that the empirical orthogonal Hankel transforms are consistent estimators of the respective orthogonal Hankel transforms. Therefore, if the samples $\mathbf{X}$ and $\mathbf{Y}$ come from the distributions orthogonally invariant in distribution, the value of the square difference $\Big(\Tilde{\mathcal{H}}_{{n_1}, \nu} (T)-\Tilde{\mathcal{H}}_{{n_2}, \nu} (T)\Big)^2$ is expected to be small, and thus the value of the test statistic is expected to be small. This is the only known two-sample test of equality in the space of symmetric positive definite matrix distributions orthogonally invariant in distribution up to date. In the univariate case, one can find the two-sample test for the equality of distributions in \cite{baringhaus2015two}. For the explicit form of the test statistic and further properties of the test, please refer to \cite{nas1}. }

The change point test was developed following the same approach in \cite{nas2}. The selection of orthogonal Hankel transforms in \cite{hadjicosta2020integral, nas2, nas1} was motivated by their favorable theoretical properties.

{It should be noted that the tests in \cite{hadjicosta2020integral} and \cite{nas1} are not directly comparable. The test in \cite{hadjicosta2020integral} is the GOF test for the Wishart distribution, while the test in \cite{nas1} is the two-sample test of equality in distribution for symmetric positive definite matrix distributions orthogonally invariant in distribution.}

\medskip

  {\section{Novel test statistic}\label{sec::teststat}

In this section, we present the test statistic of the novel test. } Let ${\mathbf{X}}=X_1, X_2, \dots, X_{n_1}$ and ${\mathbf{Y}}=Y_1, Y_2, \dots, Y_{n_2}$ be two independent random samples identically distributed as $X$ and $Y$ { respectively}, where $X$ and $Y$ are symmetric positive definite random matrices. 
Based on these samples, we  present the novel test statistic for testing the null hypothesis $$ H_0:  X \text{ and } Y \text{ are equal in distribution}.$$
 Since the Laplace transform uniquely determines the distribution, the null hypothesis can be reformulated as:
$$H_0:  \mathcal{L}_{X}(T) = \mathcal{L}_{Y}(T),\; \text{ for all } T>0, $$ 
where { $\mathcal{L}_{X} (T)=E\exp(-tr(TX))$} denotes the Laplace transform of the random variable $X$.
Since the notion of equivalence in distribution is defined in terms of the equality of corresponding Laplace transforms, a natural way to construct a test in a way to look for a difference of appropriate empirical counterparts, i.e.  empirical Laplace transforms. 
{ Let us denote the empirical Laplace transform of $\mathbf{X}$ with $\hat{L}_{n_1}(T)$, i.e.,
\begin{equation*}
    \hat{L}_{n_1} (T)=\frac{1}{n_1}\sum\limits_{k=1}^{n_1} \exp(-tr(TX_k)).
\end{equation*}
 Similarly, let us denote the empirical Laplace transform of $\mathbf{Y}$ with $\hat{L}_{n_2}(T)$, i.e., 
\begin{equation*}
    \hat{L}_{n_2} (T)=\frac{1}{n_2}\sum\limits_{k=1}^{n_2} \exp(-tr(TY_k)).
\end{equation*}
}

In the direct computations, we will utilize a noncentral Wishart measure.  It leads us to the following statistic
\begin{align}\label{dvauzorka}
    L_{n_1, n_2, \nu, \Sigma, \omega}&={ \int_{T>0}(\hat{L}_{n_1}(T)-\hat{L}_{n_2}(T))^2 dNCW(T)}\nonumber\\ &=\int_{T>0}\Big(\frac{1}{n_1}\sum\limits_{k=1}^{n_1} \exp(-tr(TX_k))-\frac{1}{n_2}\sum\limits_{k=1}^{n_2} \exp(-tr(TY_k))\Big)^2dNCW(T),
\end{align}
where $dNCW(T)$ is a corresponding noncentral Wishart { $(\nu, \Sigma, \omega)$} measure.

The noncentral Wishart distribution can be defined traditionally in the following manner: having Gaussian random vectors $Y_1, Y_2,\dots, Y_n\in \mathbb{R}^d$ with the same covariance matrix $\Sigma$ and not necessarily same means $m_1, m_2, \dots, m_n$, the vector $X=Y_1Y_1^T+Y_2Y_2^T+\dots+Y_nY_n^T$ has a noncentral Wishart distribution with the parameters $(n, \Sigma, \omega)$, where $\omega=m_1m_1^T+m_2m_2^T+\dots+m_nm_n^T$. We denote such distribution with $NCW(n, \Sigma, \omega)$. Moreover, the noncentral Wishart distribution $NCW(2\nu, \Sigma, \omega)$, where $\nu>0$ and is not neccessary an integer, exists if either $\nu \geq \frac{d-1}{2}$ or if $\nu\in \{0.5, 1, 1.5, \dots, (d-2)/2\}$ and $\text{rank}(\omega)\geq 2\nu$ \cite{letac2018laplace}. The explicit form of the density function is cumbersome and is readily available in the literature (see e.g. \cite[p. 12]{letac2004tutorial}).

The Laplace transform of the $NCW(2\nu, \Sigma, \omega)$ distribution is given by:
\begin{align}\label{laplas}
\mathcal{L}_X(s)= E(\exp(-tr(sX)))=\frac{\exp(-tr(2s(I_d+2\Sigma s)^{-1}\omega))}{det(I_d+2\Sigma s)^\nu}.
\end{align}

The direct computation, taking into account the basic identity $tr(AB)=tr(BA)$,  yields

        \begin{align*}
    L_{n_1, n_2, \nu, \omega}=\frac{1}{n_1^2n_2^2}\sum\limits_{i=1, j=1}^{n_1}\sum\limits_{l=1, k=1}^{n_2} \Psi(X_i, X_j;Y_k, Y_l),
\end{align*}
where $\Psi(X_i, X_j;Y_k, Y_l)$ is a symmetric kernel of the form: 
\begin{align*}
    \Psi(X_i, X_j;Y_k, Y_l)&=\Big(\mathcal{L}(X_i+X_j)+
    \mathcal{L}(Y_k+Y_l)-\frac{1}{2}\mathcal{L}(X_i+Y_k)-\\ 
    &\frac{1}{2}\mathcal{L}(X_i+Y_l)-\frac{1}{2}\mathcal{L}(X_j+Y_k)-\frac{1}{2}\mathcal{L}(X_j+Y_l)\Big),
\end{align*}
where the function $\mathcal{L}(\cdot)$ is { the Laplace transform of the $NCW(\nu, \Sigma, \omega)$ distribution, and is of the form given in equation} \eqref{laplas}.

{ The logic underpinning the construction of the test statistic \eqref{dvauzorka} parallels that of the test statistic \eqref{NasStari}. The empirical Laplace transforms, $\hat{L}_{n_1}(T)$ and $\hat{L}_{n_2}(T)$, are consistent estimators of the respective Laplace transforms $\mathcal{L}_X(T)$ and $\mathcal{L}_Y(T)$. Consequently, the squared difference $(\hat{L}_{n_1}(T)-\hat{L}_{n_2}(T))^2$ is expected to be small when the samples $\mathbf{X}$ and $\mathbf{Y}$ are drawn from identical distributions.  Therefore, under the null hypothesis, the value of the test statistic is expected to be small.}
 
\section{Power study}\label{sec::power}
In this section, we present the results of the power study. We focus on dimensions $d=2$ and $d=3$, while fixing the parameter $\Sigma$ to be $I_d$ for simplicity. To assess the impact of the parameters on the test power, we consider three values of $\nu$: $\nu=1$, $\nu = 2$, and $\nu=5$, and two values of $\omega$: $\omega = I_d$, and $\omega = 2I_d$.

To obtain the empirical powers, we utilize a warp-speed bootstrap algorithm with $N=10,000$ replications, similar to the approach employed in \cite{nas1}. The pseudocode for the warp-speed bootstrap algorithm is provided below (Algorithm \ref{algo}). The computation is performed using MATLAB \cite{MATLAB}. While the matrix inverse computation is the most computationally intensive operation, evaluating the test statistic is not as computationally demanding as { evaluating} the { test statistic} \eqref{NasStari}, since it does not involve special functions. However, the computational time required does increase with the dimensionality of the problem.


The level of significance is set to $\alpha=0.05$, and large values of the test statistic are considered to be significant. In all cases, we assume $d$ denotes the dimension of the respective matrices. When estimating sample covariance matrices, samples of size $d$ have been considered. The following distributions were considered:
\begin{enumerate}
    \item Wishart distribution with the shape parameter $a$ and the scale matrix $\Sigma$, denoted by $W_d(a, \Sigma)${, with a density
    \begin{equation*}
    f_{W, a, \Sigma}(X) = \frac{1}{\Gamma_d(a)}(\det\Sigma)^a(\det X)^{a-\frac{1}{2}(d+1)}\exp(tr(-\Sigma X);
    \end{equation*}
    
    }
    \item Inverse Wishart distribution with the shape parameter $a$ and the scale matrix $\Sigma$, denoted by $IW_d(a, \Sigma)$, { with a density
    \begin{equation*}
        f_{IW, a, \Sigma} (X)=\dfrac{(\det \Sigma)^{\frac{a}{2}}\exp(tr(-\frac{1}{2}\Sigma X^{-1})}{2^{\frac{a d}{2}} \Gamma_d(\frac{a}{2})(\det X)^{\frac{a+d+1}{2}}};
    \end{equation*}
    }
    \item Sample covariance matrix distributions obtained from the uniform vectors $(U_1, \dots, U_d)$, where $U_i\in \mathcal{U}[0, 1]$, denoted by $CMU_d$, { with a density
    \begin{equation*}
        f_{(U_1, \dots, U_d)} ((x_1, \dots, x_d))=1, \; x_i\in [0, 1],\; 1\leq i \leq d;
    \end{equation*}
}
 \item Sample covariance matrix distributions obtained from the random vectors having the multivariate  $t$ distribution  with $a$ degrees of freedom, denoted by $CMT_d(a, \Sigma)$, { with a density
\begin{equation*}
        f_{t, a} (x)=\frac{1}{(\det(\Sigma))^\frac{1}{2}}\frac{\Gamma(\frac{a+d}{2})}{\Gamma(\frac{d}{2})(a\pi)^\frac{d}{2}}(1+\frac{x'\Sigma^{-1}x}{a})^{-\frac{a+d}{2}}.
    \end{equation*} }
\end{enumerate}

Denote with $K_2$ the following covariance matrix:
$K_2=\begin{bmatrix}
\cos(0.7) & \sin(0.7) \\
 \sin (0.7) &\cos(0.7) 
\end{bmatrix},$
 and denote with $K_3$ the following covariance matrix:
$K_3=\begin{bmatrix}
1 & -1 & 0.95\\
-1 & 5 & 0.01\\
0.95 & 0.01 & 7
\end{bmatrix}.$

{ The results of the power study are presented in Tables \ref{pow2eye1}, \ref{pow3eye1}, \ref{pow2eye2}, \ref{pow3eye2}, \ref{pow2eye5}, and \ref{pow3eye52}. The percentages of rejected null hypotheses, obtained by the warp speed bootstrap method, serve as an estimate of the test power. This estimate is referred to as the empirical test power. The values on the diagonals represent the estimates of type I error and are relatively close to the level of significance, $\alpha=5\%$. Therefore, no significant size distortions are present for the novel test. }

Whenever a matrix in Tables \ref{pow2eye1}, \ref{pow3eye1}, \ref{pow2eye2}, \ref{pow3eye2}, \ref{pow2eye5}, and \ref{pow3eye52} is symmetric, the lower part of the table remains empty.

Based on Tables \ref{pow2eye1}, \ref{pow3eye1}, \ref{pow2eye2}, \ref{pow3eye2}, \ref{pow2eye5}, and \ref{pow3eye52}, it can be concluded that the { empirical} test powers are highly sensitive to the choice of parameters. In most cases, when the parameter $\Sigma$ is equal to $I_d$, the { empirical} test powers are higher compared to the case of $\Sigma = 2I_d $. Additionally, lower values of the parameter $\nu$ correspond to higher { empirical} test powers. The { empirical} test powers for $\nu = 5$ are generally the lowest. Given these conclusions, we recommend using the test for $\nu = 1$ and $\Sigma = I_d$ for optimal performance. Furthermore, it can be observed that the { empirical} test powers decrease as the dimension increases.

In the following, { we will compare the test outlined in \cite{nas1} and the novel test.} It is important to note that the test presented in { \cite{nas1}} tests for the orthogonal invariance in distribution, while the { novel} test tests for the equivalence in distribution.  As a result, these two tests are not directly comparable. However, one can compare the { empirical} test powers of these tests. Additionally, it is crucial to consider the theoretical properties and limitations of each test in order to make informed comparisons.

{ Given that the novel test represents the first instance of testing the equality in distribution of symmetric positive definite matrix distributions, it is not possible to make any other comparisons of empirical test powers.}

When comparing the novel test with the test { in \cite{nas1}}, it can be observed that the novel test generally outperforms the test { in \cite{nas1}}, particularly when $\nu = 1$ and $\Sigma = I_d$. The superior performance is attributed to the fact that the test { in \cite{nas1}}  focuses on orthogonal invariance in distribution rather than equivalence in distribution. However, as the parameter $\nu$ increases, the difference in { empirical} test powers between the novel test and the test { in \cite{nas1}} becomes less pronounced. It is worth noting that the novel test does not consistently outperform the test { in \cite{nas1}} in all scenarios. For instance, when testing the inverse Wishart against the Wishart distribution, the test { in \cite{nas1}} demonstrates the best performance.

We further investigate the performance of our test in a well-known theoretical scenario{, which was considered in \cite{nas1} as well}. It is important to highlight that if we have a collection of vectors $x_1, x_2, \dots, x_n$ that follow a multivariate normal distribution $N_p(0, \Sigma)$, and we represent them as a matrix $X=(x_1, x_2, \dots, x_n)$, where $n$ is greater than or equal to $p$, then $X'X$ is positive definite and belongs to the matrix-variate Wishart distribution $W_p(n, \Sigma)$ \cite[p. 88]{gupta1999matrix}.

\begin{figure}[htbp]
\caption{The case of $2\times 2$ matrices.}\label{2times2cmt}
\centering
\includegraphics[scale = 0.6]{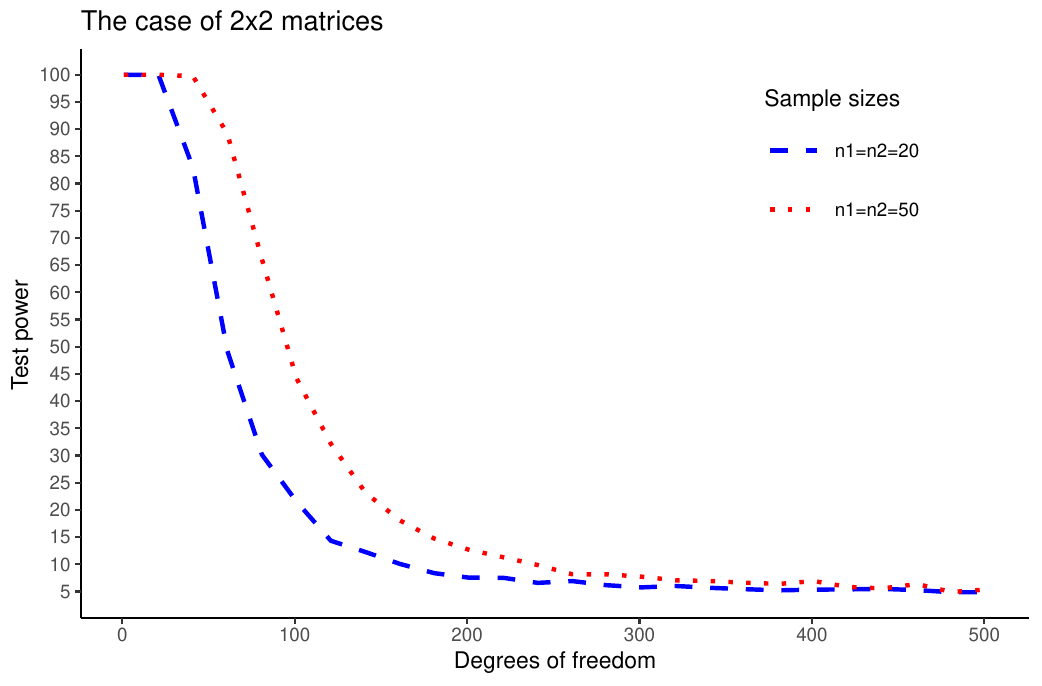}
\end{figure}

\begin{figure}[htbp]
\caption{The case of $3\times 3$ matrices.}\label{3times3cmt}
\centering
\includegraphics[scale = 0.6]{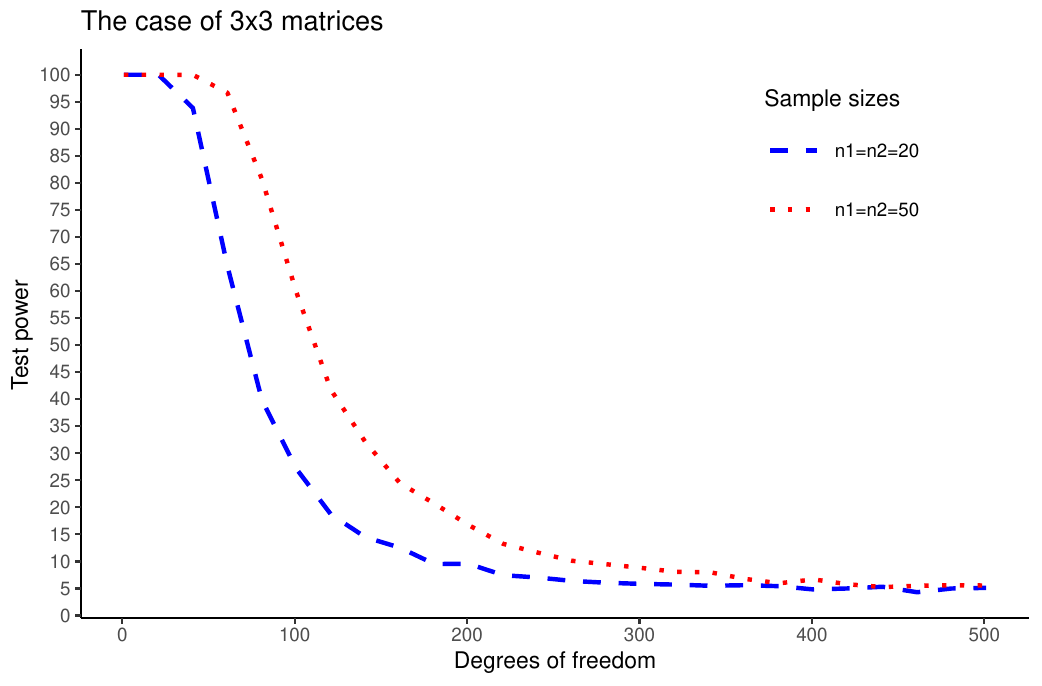}
\end{figure}

In order to assess the behavior of the novel test in this scenario, we tested for the equality in distribution of the appropriately scaled Wishart distribution $\frac{1}{499}W_d(500, I_d)$ against the $CMT_d(df, I_d)$ distribution. The sample covariance matrices were computed based on $NCov=500$ randomly generated vectors. The parameter $df$ was selected from the set ${1, 21, 41, \dots, 501}$. We fixed the parameters of the novel test statistic at $\nu=1$ and $\Sigma = I_d$ for simplicity. The results for $d=2$ and $d=3$ are presented in Figures \ref{2times2cmt} and \ref{3times3cmt} correspondingly, with the { empirical} test powers expressed as percentages.

As anticipated, our tests demonstrate the expected behavior. With an increase in the degrees of freedom, the $t$-distribution tends to approach the normal distribution. Consequently, the distribution of the covariance matrices becomes more similar to the adequately scaled Wishart distribution. This ultimately leads to a gradual decrease in { empirical} test powers, and attaining the test size for higher degrees of freedom.
\newpage

\begin{table}[htbp] 
\caption{{ The percentage of rejected $H_0$} for different sample sizes for $2\times 2$ matrices, $\nu=1, \omega = I_d (\nu=2, \omega=I_d)$.}\label{pow2eye1}
\begin{adjustbox}{width=\linewidth,center}

\begin{tabular}{l|lllllllllll}
\hline
$n_1=20, n_2=20$ & $W_{2}(2.5, I_2)$ & $IW_{2}(2.5, I_2)$ & $CMT_2(1, I_2)$ &$CMU_2$ & $W_{2}(2.5, 2I_2)$ & $IW_{2}(4, 2.5I_2)$ & $W_2(2.5, K_2)$ & $CMT_2(3, K_2)$ & $CMT_2(5, K_2)$ & $CMT_2(3, I_2)$ & $CMT_2(5, I_2)$ \\ \hline
$W_{2}(2.5, I_2)$ & 4 (4) & 22 (24) & 29 (21) & 100 (100) & 51 (39) & 20 (21) & 99 (86)  & 91 (76) & 96 (85) & 99 (91) & 100 (97) \\
$IW_{2}(2.5, I_2)$ &  & 5 (5) & 16 (9) & 100 (100) & 87 (81) & 9 (10) & 97 (63) & 71 (54) & 83 (66) & 97 (80) & 99 (88)\\
$CMT_2(1, I_2)$ &  &  & 5 (5) &100 (100)  & 60 (37) & 47 (16) & 92 (30) & 50 (34) & 64 (45) & 86 (60) & 93 (73)\\
$CMU_2$ &  &  &  & 5 (5) & 100 (100) & 100 (100)  & 100 (100)& 100 (100)& 100 (100)& 100 (100)& 100 (100)\\
$W_{2}(2.5, 2I_2)$ &  &  &  &  & 4 (4) & 95 (92) & 100 (99) & 99 (91) & 100 (96) & 100 (98) & 100 (99) \\
$IW_{2}(4, 2.5I_2)$ &  &  &  &  &  &  5 (5) & 89 (89) & 94 (68) & 89 (82) & 99 (88) & 100 (95) \\ 
$W_2(2.5, K_2)$ &  &  &  &  &  &   & 4 (4) & 83 (52) & 90 (63) & 67 (57) & 76 (71)  \\
$CMT_2(3, K_2)$ &  &  &  &  &  &   &   & 5 (5) & 6 (6) & 29 (13)  & 41 (18) \\
$CMT_2(5, K_2)$ &  &  &  &  &  &   &   &  & 5 (5) & 24 (9) & 31 (14) \\
$CMT_2(3, I_2)$ &  &  &  &  &  &   &   &  &  & 5 (5) & 6 (6) \\
$CMT_2(5, I_2)$ &  &  &  &  &  &   &   &  &  & & 5 (5)\\

\hline
\end{tabular}
\end{adjustbox}

\medskip 

\begin{adjustbox}{width=\linewidth,center}
\begin{tabular}{l|lllllllllll}
\hline
$n_1=30, n_2=20$ & $W_{2}(2.5, I_2)$ & $IW_{2}(2.5, I_2)$ & $CMT_2(1, I_2)$ &$CMU_2$& $W_{2}(2.5, 2I_2)$ & $IW_{2}(4, 2.5I_2)$ & $W_2(2.5, K_2)$ &  $CMT_2(3, K_2)$ & $CMT_2(5, K_2)$ & $CMT_2(3, I_2)$ & $CMT_2(5, I_2)$\\ \hline
$W_{2}(2.5, I_2)$ & 5 (4) & 30 (32) & 44 (30)  & 100 (100) & 58 (42) & 22 (21) & 100 (96) & 97 (89) & 99 (94) & 100 (97) & 100 (99) \\
$IW_{2}(2.5, I_2)$ & 28 (29) & 5 (5) & 32 (15) & 100 (100) & 92 (89) & 7 (8) & 99 (82) & 85 (70) & 92 (81) & 100 (91) & 100 (96) \\
$CMT_2(1, I_2)$ & 27 (16) & 12 (7) & 4 (4) & 100 (100) & 64 (39) & 48 (13) & 96 (35) & 59 (41) & 73 (55) & 94 (73) & 98 (83) \\
$CMU_2$ & 100 (100) & 100 (100) & 100 (100)& 5 (5) & 100 (100) & 100 (100)& 100  (100)& 100 (100)& 100 (100)& 100 (100)&  100 (100)\\
$W_{2}(2.5, 2I_2)$ & 65 (52) & 94 (89) & 77 (55)  & 100 (100)& 3 (4) & 98 (96) & 100 (100) & 100 (98) & 100 (99) & 100 (100)& 100 (100) \\
$IW_{2}(4, 2.5I_2)$ & 30 (29) & 13 (14) & 82 (32) & 100 (100) & 98 (97) & 5 (5) & 100 (97) & 96 (83) & 99 (92) & 100 (96) & 100 (99) \\
$W_2(2.5, K_2)$ & 100 (93) & 99 (76) & 98 (54) & 100 (100) & 100 (100) & 100 (95) & 4 (5)& 94 (67)& 96 (78)& 78 (72)& 88 (81)\\
$CMT_2(3, K_2)$ & 95 (84)& 77 (60) & 56 (38)& 100 (100)& 100 (97)& 94 (76)& 89 (56)& 5 (5) & 7 (7) & 37 (15)& 52 (25) \\
$CMT_2(5, K_2)$ & 98 (92)& 87 (73)& 72 (50)& 100 (100)& 100 (99)& 98 (88)& 93 (68)& 7 (6)& 4 (6)& 32 (12)& 40 (18) \\
$CMT_2(3, I_2)$ & 100 (96)& 99 (87) & 92 (67)& 100 (100)& 100 (100)& 100 (95)& 70 (63)& 37 (13)  & 31 (10)& 5 (5)& 6 (7)\\
$CMT_2(5, I_2)$ & 100 (99) & 100 (95) & 96 (79)& 100 (100)& 100 (100)& 100 (98)& 83 (76)& 49 (24) & 40 (16)& 7 (6) & 5 (4)\\
\hline
\end{tabular}
\end{adjustbox}

\medskip

\begin{adjustbox}{width=\linewidth,center}
\begin{tabular}{l|lllllllllll}
\hline
$n_1=50, n_2=20$ & $W_{2}(2.5, I_2)$ & $IW_{2}(2.5, I_2)$ & $CMT_2(1, I_2)$ &$CMU_2$& $W_{2}(2.5, 2I_2)$ & $IW_{2}(4, 2.5I_2)$ & $W_2(2.5, K_2)$ &  $CMT_2(3, K_2)$ & $CMT_2(5, K_2)$ & $CMT_2(3, I_2)$ & $CMT_2(5, I_2)$\\ \hline
$W_{2}(2.5, I_2)$ & 4 (4) & 37 (41) & 70 (44) & 100 (100)& 66 (49)& 26 (25)& 100 (99)& 99 (96)& 100 (99)& 100 (100)& 100 (100)\\
$IW_{2}(2.5, I_2)$ & 29 (33)& 5 (6)& 62 (29) & 100 (100)& 96 (93) & 5 (7)& 100 (94) & 94 (84)& 98 (92)& 100 (98)& 100 (99)\\
$CMT_2(1, I_2)$ & 22 (14)& 10 (5)& 5 (5)& 100 (100)& 70 (42)& 37 (10)& 99 (37)& 71 (52)& 83 (67)& 99 (84)& 100 (92)\\
$CMU_2$ & 100 (100)& 100 (100)&100 (100)& 5 (5)& 100 (100)& 100 (100)& 100 (100)& 100 (100)& 100 (100)& 100 (100)& 100 (100)\\
$W_{2}(2.5, 2I_2)$ & 77 (67)& 97 (95)& 92 (73) & 100 (100)& 4 (4)& 99 (99)& 100 (100)& 100 (100)& 100 (100)& 100 (100) & 100 (100)\\
$IW_{2}(4, 2.5I_2)$ & 42 (37)& 20 (20) & 100 (57)& 100 (100)& 100 (99)& 5 (5)& 100 (100)& 100 (95)& 100 (98)& 100 (100)& 100 (100)\\
$W_2(2.5, K_2)$ & 100 (97)& 100 (85) & 100 (78)& 100 (100)& 100 (100)& 100 (98)& 5 (4)& 98 (85) & 99 (91)& 88 (83)& 94 (90)\\
$CMT_2(3, K_2)$ & 98 (91)& 83 (64)& 65 (43)& 100 (100)& 100 (99)& 97 (86)& 93 (62)& 5 (5)& 7 (7)& 46 (20)& 62 (32)\\
$CMT_2(5, K_2)$ & 100 (96) & 93 (82)& 78 (59)& 100 (100)& 100 (100)& 100 (95)& 97 (79)& 6 (6)& 5 (5)& 41 (15)& 54 (22)\\
$CMT_2(3, I_2)$ & 100 (99)& 100 (93)& 96 (77)& 100 (100)& 100 (100)& 100 (99)& 79 (69)& 43 (14)& 36 (11)& 5 (5)& 7 (6)\\
$CMT_2(5, I_2)$ & 100 (100)& 100 (98)& 98 (87)& 100 (100)& 100 (100)& 100 (100)& 91 (85)& 58 (26)& 48 (17)& 7 (5)& 5 (5)\\\hline

\end{tabular}
\end{adjustbox}

\medskip 

\begin{adjustbox}{width=\linewidth,center}
\begin{tabular}{l|llllllllllll}
\hline
$n_1=50, n_2=30$ & $W_{2}(2.5, I_2)$ & $IW_{2}(2.5, I_2)$ & $CMT_2(1, I_2)$ &$CMU_2$ & $W_{2}(2.5, 2I_2)$  & $IW_{2}(4, 2.5I_2)$ & $W_2(2.5, K_2)$ &  $CMT_2(3, K_2)$ & $CMT_2(5, K_2)$ & $CMT_2(3, I_2)$ & $CMT_2(5, I_2)$  \\ \hline
$W_{2}(2.5, I_2)$ &   4 (4) &  45 (49)&  78 (45)& 100 (100)&  84 (69)&  36 (35)& 100 (100)& 100 (99)& 100 (100)& 100 (100)& 100 (100)\\
$IW_{2}(2.5, I_2)$ &  42 (44)&   5 (5) &  67 (27)& 100 (100)&  99 (98)&   9 (11)& 100 (99)&  98 (90)&  99 (96) & 100 (99)& 100 (100)\\
$CMT_2(1, I_2)$    &  53 (26)&  29 (10)&   5 (4)& 100 (100)&  91 (67)&  91 (29)& 100 (73)&  81 (62)&  92 (78)& 100 (93)& 100 (97)\\
$CMU_2$            & 100 (100)& 100 (100)& 100 (100)&   5 (5)& 100 (100)& 100 (100)& 100 (100)& 100 (100)& 100 (100)& 100 (100)& 100 (100)\\
$W_{2}(2.5, 2I_2)$ &  87 (77)&  99 (99)&  96 (78)& 100 (100)&   4 (4)& 100 (100)& 100 (100)& 100 (100)& 100 (100)& 100 (100)& 100 (100)\\
$IW_{2}(4, 2.5I_2)$&  49 (46)&  17 (20)& 100 (63)& 100 (100)& 100 (100)&   5 (5)& 100 (100)& 100 (98)& 100 (99)& 100 (100)& 100 (100)\\
$W_2(2.5, K_2)$    & 100 (100)& 100 (98) & 100 (91)& 100 (100)& 100 (100)& 100 (100)&   4 (4)& 100 (91)& 100 (95)&  93 (89)&  98 (96)\\
$CMT_2(3, K_2)$    & 100 (98)&  95 (85)&  79 (59)& 100 (100)& 100 (100)& 100 (96)&  99 (82)&   5 (5)&   8 (6)&  63 (24)&  78 (41)\\
$CMT_2(5, K_2)$    & 100 (99)&  99 (93)&  91 (75)& 100 (100)& 100 (100)& 100 (99)& 100 (91)&   8 (6)&   5 (5)&  57 (17)&  69 (28) \\
$CMT_2(3, I_2)$    & 100 (100)& 100 (99)&  99 (89)& 100 (100)& 100 (100)& 100 (100)&  91 (86)&  62 (21)&  53 (14)&   4 (4)&   7 (7)\\
$CMT_2(5, I_2)$    & 100 (100) & 100 (100)& 100 (96)& 100 (100)& 100 (100)& 100 (100)&  97 (95)&  74 (37)&  66 (26)&   7 (7)&   5 (5)\\
\hline

\end{tabular}
\end{adjustbox}

\medskip 

\begin{adjustbox}{width=\linewidth,center}
\begin{tabular}{l|llllllllllll}
\hline
$n_1=50, n_2=50$ & $W_{2}(2.5, I_2)$ & $IW_{2}(2.5, I_2)$ & $CMT_2(1, I_2)$ &$CMU_2$ & $W_{2}(2.5, 2I_2)$ & $IW_{2}(4, 2.5I_2)$ & $W_2(2.5, K_2)$&  $CMT_2(3, K_2)$ & $CMT_2(5, K_2)$  & $CMT_2(3, I_2)$ & $CMT_2(5, I_2)$  \\ \hline
$W_{2}(2.5, I_2)$ & 4  (4) & 54 (61) & 86  (50)& 100 (100)& 94 (88) & 56 (55) & 100 (100)& 100 (100)& 100 (100)& 100 (100)& 100 (100)\\
$IW_{2}(2.5, I_2)$ &     & 5 (5)  & 79 (26) & 100 (100)& 100 (100)& 19 (20) & 100 (100)& 99 (95) & 100 (99)& 100 (100)& 100 (100)\\
$CMT_2(1, I_2)$ &   &    & 6 (4)  & 100 (100)& 100 (85)& 100 (68)& 100 (98)& 100 (73)& 100 (87)& 100 (97) & 100 (100) \\
$CMU_2$ &  &  &  & 5 (5)& 100 (100)& 100 (100)& 100 (100)& 100 (100)& 100 (100)& 100 (100)& 100 (100)\\
$W_{2}(2.5, 2I_2)$ &  &  &  &  & 4  (4)& 100 (100)& 100 (100)& 100 (100)& 100 (100) & 100 (100)& 100 (100)\\ 
$IW_{2}(4, 2.5I_2)$ &  &  &  &  &   & 5 (4)& 100 (100)& 100 (100)& 100 (100)& 100 (100)& 100 (100)\\
$W_2(2.5, K_2)$ &  &  &  &  &   &  & 4 (5)& 100 (96)& 100 (99)& 98 (95)& 100 (99)\\
$CMT_2(3, K_2)$ &  &  &  &  &  &   &   & 6 (5)& 8 (7)& 80 (30)& 92 (52) \\
$CMT_2(5, K_2)$ &  &  &  &  &  &   &   &  & 4 (4)& 74 (21)& 85 (37)\\
$CMT_2(3, I_2)$ &  &  &  &  &  &   &   &  &  & 5 (6) & 8 (8) \\
$CMT_2(5, I_2)$ &  &  &  &  &  &   &   &  &  & & 5 (5)\\
\hline
\end{tabular}
\end{adjustbox}
\end{table}

\begin{algorithm}
  \caption{Warp-speed bootstrap algorithm}\label{algo}
  \begin{algorithmic}[1]
    \State Sample $\textbf{x}=(x_1, \dots, x_{n_1})$ from $ F_X$ and $\mathbf{y}=y_1, \dots, y_n$ from $F_Y$;
    \State Compute $L_{n_1, n_2, \nu, \Sigma, \omega}(\textbf{x}, \textbf{y})$;
    \State Generate bootstrap samples $\textbf{x*}=(x^*_1, \dots, x^*_{n_1})$  and  $\textbf{y*}=y^*_1, \dots, y^*_{n_2}$ from $F_{n_1+n_2}$- sampling distribution based on the joint sample $(\textbf{x},\textbf{y})$; 
    \State Compute $L_{n_1, n_2, \nu, \Sigma, \omega}(\textbf{x}*,\textbf{y}*)$;
    \State Repeat steps 1-4 N times and obtain two sequences of statistics $\{L_{n_1, n_2, \nu, \Sigma, \omega}^{(j)}\}$ and $\{L_{n_1, n_2, \nu, \Sigma, \omega}^{*(j)}\}$,  $j=1,...,N$; 
    \State Reject the null hypothesis for the $j$--sample ($j=1,...,N$), if $L_{n_1, n_2, \nu, \Sigma, \omega}^{(j)}>c_\alpha$,  where $c_\alpha$ denotes the $(1-\alpha)\%$ quantile of the empirical distribution of the bootstrap test statistics  $(L_{n_1, n_2, \nu, \Sigma, \omega}^{*(j)}, \ j=1,...,N)$.
     
  \end{algorithmic}
\end{algorithm}
\newpage

\begin{table}[htbp]
\caption{{ The percentage of rejected $H_0$} for different sample sizes for $3\times 3$ matrices,  $\nu=1, \omega = I_d (\nu=2, \omega=I_d)$.}\label{pow3eye1}
\begin{adjustbox}{width=\linewidth,center}
\begin{tabular}{l|lllllllllllll}
\hline
$n_1=20, n_2=20$ & $W_3(3, I_3)$ & $IW_3(3, I_3)$ & $CMT_3(1, I_3)$ &$CMU_3$& $W_3(3, 2I_3)$ & $IW_3(5, 3I_3)$ & $W_3(3, K_3)$ & $CMT_3(3, K_3)$ & $CMT_3(5, K_3)$  & $CMT_3(3, I_3)$ & $CMT_3(5, I_3)$\\ \hline
$W_3(3, I_3)$ & 3 (3) & 14 (18)& 12 (4)& 100 (100)& 59 (29)& 40 (36)& 99 (63)& 83 (58)& 95 (74)& 92 (67)& 99 (81)\\
$IW_3(3, I_3)$ &   & 5 (4) & 27 (13)& 100 (100) & 73 (56)& 11 (6) & 95 (67) & 58 (35) & 79 (50) & 73 (43)& 90 (60)\\
$CMT_3(1, I_3)$ &  &    & 3 (3) & 100 (100)& 14 (9)& 82 (23) & 37 (13) & 88 (53)& 96 (71)& 94 (62)& 99 (78) \\
$CMU_3$ &   &    &    & 5  (5) & 100 (100) & 100 (100)& 100 (100)& 100 (100)& 100 (100)& 100 (100)& 100 (100)\\
$W_3(3, 2I_3)$ &   &    &    &     & 3 (3) & 100 (98)& 67 (13) & 99 (76)& 100 (90)& 100 (80)& 100 (94)\\
$IW_3(5, 3I_3)$ &     &    &    &     &    & 5 (4) & 100 (100)& 67 (43) & 85 (59)& 82 (52)& 94 (69) \\
$W_3(3, K_3)$   &    &    &     &    &    & & 3 (3)  & 100 (82)& 100 (93)& 100 (85)& 100 (95) \\
$CMT_3(3, K_3)$ &     &    &    &     &    &    &     & 4 (4)  & 6 (5)  & 9  (6) & 16 (9)\\
$CMT_3(5, K_3)$ &     &    &    &     &    &    &     &     & 4  (3) & 9 (4)  & 10 (5)\\
$CMT_3(3, I_3)$ &      &    &    &     &    &    &     &     &     & 4 (4)  & 6 (5)\\
$CMT_3(5, I_3)$ &      &   &    &     &    &    &     &     &     &     & 4 (5) \\
\hline
\end{tabular}
\end{adjustbox}

\medskip 

\begin{adjustbox}{width=\linewidth,center}
\begin{tabular}{l|lllllllllll}
\hline
$n_1=30, n_2=20$ & $W_3(3, I_3)$ & $IW_3(3, I_3)$ & $CMT_3(1, I_3)$ &$CMU_3$& $W_3(3, 2I_3)$ & $IW_3(5, 3I_3)$ & $W_3(3, K_3)$ & $CMT_3(3, K_3)$ & $CMT_3(5, K_3)$ & $CMT_3(3, I_3)$ & $CMT_3(5, I_3)$\\ \hline
$W_3(3, I_3)$ & 3  (3) & 21 (27)& 27 (7) & 100 (100)& 68 (29)& 44 (43) & 100 (68)& 94 (77)& 99 (89)& 98 (82)& 100 (94)\\
$IW_3(3, I_3)$ & 15 (17) & 4 (4) & 38 (16)& 100 (100)& 82 (61)& 7 (4) & 98 (77) & 73 (51)& 90 (70)& 88 (60)& 97 (78) \\
$CMT_3(1, I_3)$ & 13 (3) & 33 (16) & 3 (3) & 100 (100)& 10 (6)& 87 (24)& 32 (10) & 95 (72) & 99 (87)& 98 (82)& 100 (93)\\
$CMU_3$ & 100 (100)& 100 (100)& 100 (100)& 4 (5) & 100 (100)& 100 (100)& 100 (100)& 100 (100)& 100 (100)& 100 (100) & 100 (100)\\
$W_3(3, 2I_3)$ &78 (52) & 85 (73)& 30 (19) & 100 (100) & 3 (3) & 100 (99) & 77 (9) & 100 (92)& 100 (98) & 100 (95)& 100 (99) \\
$IW_3(5, 3I_3)$ & 55 (48) & 19 (11) & 97 (50) & 100 (100)& 100 (99)& 4 (5) & 100 (100)& 85 (62)  & 95 (78)  & 95 (71)  & 99 (86)\\
$W_3(3, K_3)$ & 100 (87) & 99 (84) & 59 (24) & 100 (100)& 87 (24)& 100 (100)& 3 (4)   & 100 (94) & 100 (99)& 100 (96) & 100 (100) \\
$CMT_3(3, K_3)$ & 90 (62) & 60 (35) & 93 (59) & 100 (100)& 100 (85) & 70 (42)  & 100 (89) & 3 (4)  & 8 (6)  & 10 (6)  & 22 (12) \\
$CMT_3(5, K_3)$ & 98 (83) & 82 (55)& 98 (79)& 100 (100)& 100 (88)& 90 (53) & 100 (94)& 6 (5)  & 4 (5)  & 11 (4)  & 15 (6)\\
$CMT_3(3, I_3)$ & 97 (70)  & 78 (45) & 97 (71)& 100 (100) & 100 (88) & 87 (53)  & 100 (94) & 10 (5) & 12 (5)  & 4  (4) & 8 (6) \\
$CMT_3(5, I_3)$ & 100 (89) & 94 (66)& 99 (87)& 100 (100)& 100 (97) & 98 (74)  & 100 (99)& 19 (8) & 12 (5)  & 6 (5)  & 4 (4) \\

\hline
\end{tabular}
\end{adjustbox}

\medskip

\begin{adjustbox}{width=\linewidth,center}
\begin{tabular}{l|lllllllllll}
\hline
$n_1=50, n_2=20$ & $W_3(3, I_3)$ & $IW_3(3, I_3)$ & $CMT_3(1, I_3)$ &$CMU_3$ & $W_3(3, 2I_3)$  & $IW_3(5, 3I_3)$ & $W_3(3, K_3)$ & $CMT_3(3, K_3)$ & $CMT_3(5, K_3)$ & $CMT_3(3, I_3)$ & $CMT_3(5, I_3)$ \\ \hline
$W_3(3, I_3)$ & 4 (4) & 30 (37)& 56 (13)& 100 (100)& 77 (32)& 51 (50)& 100 (83) & 99 (91)& 100 (98) & 100 (96) & 100 (99)\\
$IW_3(3, I_3)$ & 14 (15) & 5 (5)& 51 (22) & 100 (100)& 89 (69)& 6 (4)& 100 (84) & 89 (71) & 97 (85) & 97 (78)& 100 (92)\\

$CMT_3(1, I_3)$ & 12 (3) & 39 (19) & 3 (4) & 100 (100)& 7 (3) & 92 (26)& 32 (6)& 99 (87) & 100 (96) & 100 (93) & 100 (99)\\

$CMU_3$ & 100 (100)& 100 (100)& 100 (100)& 5 (5)& 100 (100)& 100 (100)& 100 (100)& 100 (100)& 100 (100)& 100 (100)& 100 (100)\\

$W_3(3, 2I_3)$ & 91 (76)& 94 (87)& 54 (33)& 100 (100)& 3 (4)& 100 (100)& 89 (8)& 100 (98)& 100 (100)& 100 (99)& 100 (100)\\

$IW_3(5, 3I_3)$ & 70 (63)& 35 (16)& 100 (90)& 100 (100)& 100 (100)& 5 (4)& 100 (100)& 96 (82)& 99 (93)& 99 (88)& 100 (97)\\
 
$W_3(3, K_3)$ & 100 (98) & 100 (95) & 83 (43) & 100 (100)& 96 (44) & 100 (100)& 3 (4)& 100 (99) & 100 (100)& 100 (100)& 100 (100)\\
 
$CMT_3(3, K_3)$ & 94 (68)& 68  (33) & 97 (70)& 100 (100)& 100 (92)& 73 (41)& 100 (96)& 4 (4)& 9 (8)& 15 (8)& 31 (17)\\

$CMT_3(5, K_3)$ & 100 (89)& 90 (61)& 99 (88) & 100 (100)& 100 (96)& 93 (67)& 100 (100)& 6 (5)& 5 (4)& 14 (5) & 18 (9)\\

$CMT_3(3, I_3)$ & 99 (82)& 86 (47)& 99 (78)& 100 (100)& 100 (96)& 94 (59)& 100 (97) & 12 (5) & 15 (6)& 4 (4)& 10 (8) \\

$CMT_3(5, I_3)$ & 100 (95)& 97 (75)& 100 (93)& 100 (100)& 100 (100)& 99 (84)& 100 (100)& 26 (9)& 16 (6)& 7 (4)& 5 (5)\\

\hline
\end{tabular}
\end{adjustbox}

\medskip 

\begin{adjustbox}{width=\linewidth,center}
\begin{tabular}{l|lllllllllll}
\hline
$n_1=50, n_2=30$ & $W_3(3, I_3)$ & $IW_3(3, I_3)$ & $CMT_3(1, I_3)$ &$CMU_3$ & $W_3(3, 2I_3)$  & $IW_3(5, 3I_3)$ & $W_3(3, K_3)$ & $CMT_3(3, K_3)$ & $CMT_3(5, K_3)$ & $CMT_3(3, I_3)$ & $CMT_3(5, I_3)$  \\ \hline
$W_3(3, I_3)$ & 3 (3)& 36 (42)& 69 (10)& 100 (100)& 93 (61)& 68 (71)& 100 (98)& 100 (96)& 100 (99)& 100 (98)& 100 (100)\\
$IW_3(3, I_3)$ & 24 (28) & 5 (3) & 63 (30)& 100 (100)& 97 (86)& 14 (4)& 100 (96)& 93 (75)  & 99 (91) & 99 (85) & 100 (96)\\
$CMT_3(1, I_3)$ &34 (4) & 55 (26)& 4 (3) & 100 (100)& 24 (9)& 99 (60) & 65 (16) & 100 (94)& 100 (99)& 100 (97)& 100 (99)\\
$CMU_3$ & 100 (100)& 100 (100)& 100 (100)& 5 (5)& 100 (100)& 100 (100)& 100 (100)& 100 (100)& 100 (100)& 100 (100)& 100 (100)\\
$W_3(3, 2I_3)$ & 96 (82) & 98 (92)& 50 (30)& 100 (100)& 3 (3)  & 100 (100)& 99 (24) & 100 (100)& 100 (100)& 100 (100)& 100 (100)\\
$IW_3(5, 3I_3)$ &81 (77) & 34 (14) & 100 (93)& 100 (100)& 100 (100)& 5 (4) & 100 (100)& 98 (86) & 100 (96)& 100 (93) & 100 (99) \\
$W_3(3, K_3)$ & 100 (99)& 100 (98)& 87 (41)& 100 (100)& 100 (53)& 100 (100)& 3 (3) & 100 (100)& 100 (100)& 100 (100)& 100 (100)\\
$CMT_3(3, K_3)$ & 99 (90) & 86 (57) & 99 (88)& 100 (100)& 100 (99)& 93 (72) & 100 (100)& 4 (4)  & 9 (8)  & 18 (9) & 40 (20) \\
$CMT_3(5, K_3)$ & 100 (99)& 97 (80) & 100 (97)& 100 (100)& 100 (100)& 99 (90) & 100 (100)& 8 (6)  & 4 (4)  & 20 (7)  & 27 (11)  \\
$CMT_3(3, I_3)$ & 100 (95) & 97 (74)& 100 (93)& 100 (100)& 100 (100)& 99 (84) & 100 (100)& 19 (6) & 20 (7) & 4 (4)  & 10 (8) \\
$CMT_3(5, I_3)$ & 100 (99)& 100 (92)& 100 (99)& 100 (100)& 100 (100)& 100 (97)& 100 (100)& 35 (14) & 24 (8) & 8 (6)  & 4 (4)  \\

\hline
\end{tabular}
\end{adjustbox}

\medskip 

\begin{adjustbox}{width=\linewidth,center}
\begin{tabular}{l|llllllllllll}
\hline
$n_1=50, n_2=50$ & $W_3(3, I_3)$ & $IW_3(3, I_3)$ & $CMT_3(1, I_3)$ &$CMU_3$ & $W_3(3, 2I_3)$ & $IW_3(5, 3I_3)$ & $W_3(3, K_3)$ & $CMT_3(3, K_3)$ & $CMT_3(5, K_3)$ & $CMT_3(3, I_3)$ & $CMT_3(5, I_3)$\\ \hline
$W_3(3, I_3)$ & 3 (4) & 41 (47)& 82 (6)& 100 (100)& 99 (91)& 88 (87)& 100 (100)& 100 (99)& 100 (100)& 100 (100)& 100 (100)\\
$IW_3(3, I_3)$ &  & 4 (4)& 76 (42)& 100 (100)& 100 (97)& 38 (11) & 100 (100)& 97 (83)& 100 (96)& 100 (92)& 100 (99)\\
$CMT_3(1, I_3)$ &   &  & 3 (3)& 100 (100)& 55 (25)& 100 (95)& 94 (42)& 100 (98)& 100 (100)& 100 (99)& 100 (100)\\
$CMU_3$ &  &  &  & 6 (5) & 100 (100) & 100 (100)& 100 (100)& 100 (100)& 100 (100)& 100 (100)& 100 (100)\\
$W_3(3, 2I_3)$ &  & & &  &  3 (3)& 100 (100)& 100 (60)& 100 (100)& 100 (100)& 100 (100)& 100 (100)\\
$IW_3(5, 3I_3)$ &  &  &  &  &  & 5 (5)& 100 (100)& 100 (93)& 100 (99)& 100 (98)& 100 (100) \\
$W_3(3, K_3)$ &  &  &  &  &  &    & 3 (3) & 100 (100)& 100 (100)& 100 (100)& 100 (100)\\ 
$CMT_3(3, K_3)$ &  &  &  &  &  &   &   & 5 (4)& 10 (8)& 28 (9)& 55 (24) \\
$CMT_3(5, K_3)$ &  &  &  &  &  &   &   &  & 5 (4)& 30 (9)& 39 (13) \\
$CMT_3(3, I_3)$ &  &  &  &  &  &   &   &  &  & 5 (4)& 11 (9)\\
$CMT_3(5, I_3)$ &  &  &  &  &  &   &   &  &  &  & 6 (4)\\

\hline
\end{tabular}
\end{adjustbox}
\end{table}

\newpage
\begin{table}[htbp] 
\caption{{ The percentage of rejected $H_0$} for different sample sizes for $2\times 2$ matrices, $\nu=1, \omega = 2I_d (\nu=2, \omega= 2I_d)$.}\label{pow2eye2}
\begin{adjustbox}{width=\linewidth,center}

\begin{tabular}{l|lllllllllll}
\hline
$n_1=20, n_2=20$ & $W_{2}(2.5, I_2)$ & $IW_{2}(2.5, I_2)$ & $CMT_2(1, I_2)$ &$CMU_2$ & $W_{2}(2.5, 2I_2)$ & $IW_{2}(4, 2.5I_2)$ & $W_2(2.5, K_2)$ & $CMT_2(3, K_2)$ & $CMT_2(5, K_2)$ & $CMT_2(3, I_2)$ & $CMT_2(5, I_2)$ \\ \hline
$W_{2}(2.5, I_2)$ & 4 (4) & 19 (18) & 28 (19) & 100 (100) & 35 (23) & 13 (11) & 96 (67)  & 83 (62) & 90 (73) & 96 (79) & 99 (88) \\
$IW_{2}(2.5, I_2)$ &  & 5 (5) & 16 (11) & 100 (100) & 81 (73) & 11 (13) & 92 (44) & 69 (48) & 79 (59) & 91 (69) & 96 (81)\\
$CMT_2(1, I_2)$ &  &  & 4 (4) & 100 (100) & 47 (28) & 30 (17) & 60 (11) & 33 (22) & 44 (30) & 67 (45) & 76 (54) \\
$CMU_2$ &  &  &  & 5 (5) & 100 (100)  & 100  (100)  & 100 (100)& 100 (100)& 100 (99) & 100 (99) & 100 (98) \\
$W_{2}(2.5, 2I_2)$ &  &  &  &  & 4 (4) & 92 (86) & 100 (95)  & 95 (73) & 99 (84) & 100 (89) & 100 (93)  \\
$IW_{2}(4, 2.5I_2)$ &  &  &  &  &  &  5 (4)  & 99 (66) & 82 (61) & 91 (69) & 97 (78) & 99 (89) \\ 
$W_2(2.5, K_2)$ &  &  &  &  &  &   & 4 (5) & 64 (40) & 76 (49) & 62 (55) & 72 (63)  \\
$CMT_2(3, K_2)$ &  &  &  &  &  &   &   & 5 (4)  & 6 (5)   & 16 (11) & 25 (15) \\
$CMT_2(5, K_2)$ &  &  &  &  &  &   &   &  & 5 (4) & 13 (8) & 18 (12) \\
$CMT_2(3, I_2)$ &  &  &  &  &  &   &   &  &  & 5 (5)  & 6 (5) \\
$CMT_2(5, I_2)$ &  &  &  &  &  &   &   &  &  & & 5 (5) \\

\hline
\end{tabular}
\end{adjustbox}

\medskip 

\begin{adjustbox}{width=\linewidth,center}
\begin{tabular}{l|lllllllllll}
\hline
$n_1=30, n_2=20$ & $W_{2}(2.5, I_2)$ & $IW_{2}(2.5, I_2)$ & $CMT_2(1, I_2)$ &$CMU_2$& $W_{2}(2.5, 2I_2)$ & $IW_{2}(4, 2.5I_2)$ & $W_2(2.5, K_2)$ &  $CMT_2(3, K_2)$ & $CMT_2(5, K_2)$ & $CMT_2(3, I_2)$ & $CMT_2(5, I_2)$\\ \hline
$W_{2}(2.5, I_2)$ 4 (4) & 27 (27)  & 41 (29)   & 100 (100) & 38 (24) & 14 (12) & 99 (84) & 93 (79) & 97 (87) & 99 (92) & 100 (96)  \\
$IW_{2}(2.5, I_2)$ & 25 (22) & 5 (5) & 29 (18) & 100 (100) & 87 (81) & 9 (10) & 98 (69) & 83 (64) & 91 (75) & 98 (84) & 100 (91)  \\
$CMT_2(1, I_2)$ & 27 (15) & 12 (8)  & 4 (4)  & 100 (100)  & 52 (26) & 26 (16) & 66 (10) & 42 (29) & 56 (39) & 80 (56) & 90 (68) \\
$CMU_2$ & 100 (100) & 100 (100)  & 100 (100) & 5 (5)   & 100 (100)  & 100 (100) & 100 (100) & 100 (100) & 100 (100) & 100 (100) &  100 (100) \\
$W_{2}(2.5, 2I_2)$ & 50 (34) & 90 (83) & 68 (42) & 100 (100)& 3 (4)  & 95 (93) & 100 (99)  & 99 (90) & 100 (95)   & 100 (96) & 100 (99)  \\
$IW_{2}(4, 2.5I_2)$ & 23 (20) & 19 (18) & 51 (30) & 100 (100)  & 97 (94) & 5 (5)  & 100 (90)  & 93 (77) & 97 (85) & 100  (91) & 100 (95)  \\
$W_2(2.5, K_2)$ & 99 (79) & 97 (52)& 83 (21) & 100 (100) & 100 (98)  & 100 (75)  & 5 (5)& 83 (57)& 89 (66)& 75 (68)& 85 (78)\\
$CMT_2(3, K_2)$ & 88  (69) & 73 (52) & 37 (22) & 100 (100) & 99 (84) & 89 (66) & 74 (42) &  5 (5) &  6 (6) & 22 (13)  & 32 (21) \\
$CMT_2(5, K_2)$ & 95 (77) & 84 (63) & 48 (34) & 100 (100)& 100 (91)  & 95 (76) & 82 (55) &  5 (5) &  5 (5) & 17 (9) & 25 (15)\\
$CMT_2(3, I_2)$ & 99 (86) & 96 (78) & 74 (50)& 100 (100)& 100 (94) & 99 (86) & 69 (58) & 19 (11) & 14 (7) &  5 (5) &  6 (6)\\
$CMT_2(5, I_2)$ & 100 (94) & 99 (87) & 85 (63) & 100 (100)& 100 (97)& 100  (94)& 81 (69) & 29 (18) & 22 (11) &  5 (6) &  5 (5) \\
\hline
\end{tabular}
\end{adjustbox}

\medskip

\begin{adjustbox}{width=\linewidth,center}
\begin{tabular}{l|lllllllllll}
\hline
$n_1=50, n_2=20$ & $W_{2}(2.5, I_2)$ & $IW_{2}(2.5, I_2)$ & $CMT_2(1, I_2)$ &$CMU_2$& $W_{2}(2.5, 2I_2)$ & $IW_{2}(4, 2.5I_2)$ & $W_2(2.5, K_2)$ &  $CMT_2(3, K_2)$ & $CMT_2(5, K_2)$ & $CMT_2(3, I_2)$ & $CMT_2(5, I_2)$\\ \hline
$W_{2}(2.5, I_2)$ &   4 (4) &  32 (34) &  59 (43) & 100 (100)&  44 (24) &  13 (12)& 100 (96)&  99 (91)& 100 (96)& 100 (98)& 100 (99)\\
$IW_{2}(2.5, I_2)$ &  30 (28)&   5 (5)&  52 (32)& 100 (100)&  95 (87)&   7 (9)& 100 (86)&  94 (82)&  98 (87) & 100 (95) & 100 (98)\\
$CMT_2(1, I_2)$ &  22 (12)&   9 (6)&   5 (4)& 100 (100)&  55 (26)&  25 (12)&  78 (8)&  52 (37)&  68 (49)&  91 (70)&  97 (82)\\
$CMU_2$ & 100 (100)& 100 (100)& 100 (100)&   5 (5)& 100 (100)& 100 (100)& 100 (100)& 100 (100)& 100 (100)& 100 (100)& 100 (100)\\
$W_{2}(2.5, 2I_2)$ &  63 (52)&  96 (91)&  86 (62)& 100 (100)&   4 (4)&  98 (97)& 100 (100)& 100 (97)& 100 (99)& 100 (99)& 100 (100)\\
$IW_{2}(4, 2.5I_2)$  &  40 (32)&  25 (26)&  88 (49)& 100 (100)&  99 (98)&   5 (5)& 100 (99)&  99 (90)& 100 (95)& 100 (98)& 100 (99)\\
$W_2(2.5, K_2)$ & 100 (86)&  99 (57)&  96 (45)& 100 (100)& 100 (100)& 100 (87)&   4 (5)&  94 (74)&  96 (82)&  88 (81)&  94 (88)\\
$CMT_2(3, K_2)$ &  95 (79)&  79 (56)&  43 (27)& 100 (100)& 100 (92)&  94 (74)&  79 (46)&   5 (5)&   6 (5)&  28 (17) &  42 (26)\\
$CMT_2(5, K_2)$ &  98 (88) &  91 (71)&  57 (39)& 100 (100)& 100 (97)&  98 (86)&  88 (60)&   5 (4)&   5 (5)&  23 (13)&  32 (18)\\
$CMT_2(3, I_2)$ & 100 (94)&  99 (87)&  83 (55)& 100 (100)& 100 (99)& 100 (94)&  76 (64)&  23 (10)&  16 (8)&   5 (5)&   6 (7)\\
$CMT_2(5, I_2)$ & 100 (98)& 100 (94)&  92 (71)& 100 (100)& 100 (100)& 100 (98)&  85 (77)&  32 (17)&  27 (12)&   5 (6) &   5 (5) \\\hline
\end{tabular}
\end{adjustbox}

\medskip 

\begin{adjustbox}{width=\linewidth,center}
\begin{tabular}{l|llllllllllll}
\hline
$n_1=50, n_2=30$ & $W_{2}(2.5, I_2)$ & $IW_{2}(2.5, I_2)$ & $CMT_2(1, I_2)$ &$CMU_2$ & $W_{2}(2.5, 2I_2)$  & $IW_{2}(4, 2.5I_2)$ & $W_2(2.5, K_2)$ &  $CMT_2(3, K_2)$ & $CMT_2(5, K_2)$ & $CMT_2(3, I_2)$ & $CMT_2(5, I_2)$  \\ \hline
$W_{2}(2.5, I_2)$ &  4 (4)&  40 (43)&  65 (45)& 100 (100)&  64 (45)&  23 (20)& 100 (99)& 100 (95)& 100 (98)& 100 (100)& 100 (100)\\
$IW_{2}(2.5, I_2)$ &  38 (39)&   5 (5)&  56 (29)& 100 (100)&  99 (97)&  15 (15)& 100 (95)&  97 (87)&  99 (94)& 100 (98)& 100 (99)\\
$CMT_2(1, I_2)$    &  43 (23)&  23 (14)&   5 (4)& 100 (100)&  83 (47)&  55 (26)&  95 (19)&  63 (43)&  76 (59)&  97 (80) &  99 (90) \\
$CMU_2$            & 100 (100)& 100 (100) & 100 (100)&   5 (4)& 100 (100)& 100 (100)& 100 (100)& 100 (100)& 100 (100)& 100 (100)& 100 (100)\\
$W_{2}(2.5, 2I_2)$  &  74 (58)&  98 (97)&  91 (64)& 100 (100) &   4 (4)& 100 (99)& 100 (100)& 100 (99)& 100 (100)& 100 (100)& 100 (100)\\
$IW_{2}(4, 2.5I_2)$ &  43 (35)&  26 (26)&  93 (52)& 100 (100)& 100 (100)&   5 (6) & 100 (100)& 100 (96)& 100 (98)& 100 (99)& 100 (100) \\
$W_2(2.5, K_2)$    & 100 (98)& 100 (89)&  99 (45)& 100 (100)& 100 (100)& 100 (99)&   4 (4)&  97 (79)&  99 (89)&  93 (88) &  97 (94)\\
$CMT_2(3, K_2)$    &  99 (92)&  94 (78)&  58 (37)& 100 (100)& 100 (99)&  99 (91)&  95 (69)&   5 (5)&   6 (6)&  39 (19)&  57 (31)\\
$CMT_2(5, K_2)$    & 100 (97)&  98 (88)&  72 (51)& 100 (100)& 100 (100)& 100 (97)&  98 (80)&   6 (5)&   5 (5)&  30 (13)&  43 (21)\\
$CMT_2(3, I_2)$    & 100 (99)& 100 (97)&  95 (75)& 100 (100)& 100 (100)& 100 (99)&  89 (83)&  33 (16)&  28 (11)&   5 (5)&   7 (6)\\
$CMT_2(5, I_2)$    & 100 (100)& 100 (99)&  98 (87)& 100 (100)& 100 (100)& 100 (100)&  96 (91)&  49 (26)&  38 (18)&   6 (6)&   5 (5)\\
\hline

\end{tabular}
\end{adjustbox}

\medskip 

\begin{adjustbox}{width=\linewidth,center}
\begin{tabular}{l|llllllllllll}
\hline
$n_1=50, n_2=50$ & $W_{2}(2.5, I_2)$ & $IW_{2}(2.5, I_2)$ & $CMT_2(1, I_2)$ &$CMU_2$ & $W_{2}(2.5, 2I_2)$ & $IW_{2}(4, 2.5I_2)$ & $W_2(2.5, K_2)$&  $CMT_2(3, K_2)$ & $CMT_2(5, K_2)$  & $CMT_2(3, I_2)$ & $CMT_2(5, I_2)$  \\ \hline
$W_{2}(2.5, I_2)$ & 4 (4)&  55 (54)&  72 (46)& 100 (100)&  86 (70)&  46 (40)& 100 (100)& 100 (99)& 100 (100)& 100 (100)& 100 (100)\\
$IW_{2}(2.5, I_2)$  &   &   5 (4)&  56 (29)& 100 (100)& 100 (99)&  27 (26)& 100 (99)&  99 (94)& 100 (98)& 100 (100)& 100 (100)\\
$CMT_2(1, I_2)$ &   &   &   5 (5)& 100 (100)&  96 (75)&  96 (53)& 100 (58)&  73 (56)&  88 (69)&  99 (90)& 100 (96)\\
$CMU_2$ &   &   &   &   5 (5)& 100 (100)& 100 (100)& 100 (100)& 100 (100)& 100 (100)& 100 (100)& 100 (100)\\
$W_{2}(2.5, 2I_2)$  &   &   &   &   &   4 (4)& 100 (100)& 100 (100)& 100 (100)& 100 (100)& 100 (100)& 100 (100)\\ 
$IW_{2}(4, 2.5I_2)$ &   &   &   &   &   &   5 (5)& 100 (100)& 100 (99)& 100 (100)& 100 (100)& 100 (100)\\
$W_2(2.5, K_2)$     &   &   &   &   &   &   &   5 (5)&  99 (89)& 100 (94)&  98 (94)&  99 (98) \\
$CMT_2(3, K_2)$ &   &   &   &   &   &   &   &   5 (5)&   7 (6)&  50 (23)&  67 (41)\\
$CMT_2(5, K_2)$ &   &   &   &   &   &   &   &   &   5 (5)&  43 (16)&  58 (27)\\
$CMT_2(3, I_2)$ &   &   &   &   &   &   &   &   &   &   5 (4)&   7 (6)\\
$CMT_2(5, I_2)$ &  &  &  &  &  &   &   &  &  & & 4 (5)\\
\hline
\end{tabular}
\end{adjustbox}
\end{table}


\newpage

\begin{table}[htbp]
\caption{{ The percentage of rejected $H_0$} for different sample sizes for $3\times 3$ matrices,  $\nu=1, \omega = 2I_d (\nu=2, \omega=2I_d)$.}\label{pow3eye2}
\begin{adjustbox}{width=\linewidth,center}
\begin{tabular}{l|lllllllllllll}
\hline
$n_1=20, n_2=20$ & $W_3(3, I_3)$ & $IW_3(3, I_3)$ & $CMT_3(1, I_3)$ &$CMU_3$& $W_3(3, 2I_3)$ & $IW_3(5, 3I_3)$ & $W_3(3, K_3)$ & $CMT_3(3, K_3)$ & $CMT_3(5, K_3)$  & $CMT_3(3, I_3)$ & $CMT_3(5, I_3)$\\ \hline
$W_3(3, I_3)$ &    3 (3)&    14 (14)&     4 (5)&    100 (100)&    38 (13)&    30 (16)&    74 (25)&    62 (34)&    78 (44)&    74 (39)&    88 (50) \\
$IW_3(3, I_3)$ &    &     4 (4)&     9 (5)&    100 (100)&    60 (39)&     7 (7)&    84 (46)&    41 (22)&    57 (31)&    54 (28)&    73 (39)\\
$CMT_3(1, I_3)$ &    &    &     3 (4)&    100 (100)&    11 (8)&    17 (5)&    17 (8)&    56 (27)&    75 (40)&    68 (34)&    85 (44)\\
$CMU_3$ &    &    &    &     5 (5)&    100 (100)&    100 (100)&    100 (100)&    100 (100)&    100 (100)&    100 (100)&    100 (100)\\
$W_3(3, 2I_3)$  &    &    &    &    &     3 (3)&    98 (89)&    31 (6)&    84 (41)&    94 (54)&    89 (46)&    98 (61)\\
$IW_3(5, 3I_3)$  &    &    &    &    &    &     4 (4)&    100 (98)&    51 (28)&    68 (37)&    64 (31)&    81 (44)\\
$W_3(3, K_3)$   &    &    &    &    &    &    &     3 (4)&    89 (42)&    96 (55)&    91 (48)&    98 (60)\\
$CMT_3(3, K_3)$ &    &    &    &    &    &    &    &     3 (4)&     4 (4)&     5 (4)&    10 (7)\\
$CMT_3(5, K_3)$ &    &    &    &    &    &    &    &    &     3 (4)&     6 (4)&     6 (4)\\
$CMT_3(3, I_3)$ &    &    &    &    &    &    &    &    &    &     3 (4)&     4 (4)\\
$CMT_3(5, I_3)$ &      &   &    &     &    &    &     &     &     &     & 3 (4)\\
\hline
\end{tabular}
\end{adjustbox}

\medskip 

\begin{adjustbox}{width=\linewidth,center}
\begin{tabular}{l|lllllllllll}
\hline
$n_1=30, n_2=20$ & $W_3(3, I_3)$ & $IW_3(3, I_3)$ & $CMT_3(1, I_3)$ &$CMU_3$& $W_3(3, 2I_3)$ & $IW_3(5, 3I_3)$ & $W_3(3, K_3)$ & $CMT_3(3, K_3)$ & $CMT_3(5, K_3)$ & $CMT_3(3, I_3)$ & $CMT_3(5, I_3)$ \\ \hline
$W_3(3, I_3)$  & 3 (3)&  21 (20)&   9 (8)& 100 (100)&  27 (9)&  21 (20)&  82 (19)&  82 (54)&  93 (66)&  91 (59)&  97 (73)\\
$IW_3(3, I_3)$  &  15 (12)&   4 (4)&  16 (8)& 100 (100)&  71 (41)&   5 (4)&  89 (49)&  61 (40)&  80 (49)&  77 (45)&  90 (59)\\
$CMT_3(1, I_3)$ &   3 (3)&  10 (5)&   3 (4)& 100 (100)&   7 (4)&  14 (3)&  12 (5)&  77 (45)&  90 (59)&  85 (51)&  96 (66)\\
$CMU_3$ & 100 & 100 (100)& 100 (100)&   5 (5)& 100 (100)& 100 (100)& 100 (100)& 100 (100)& 100 (100)& 100 (100)& 100 (100)\\
$W_3(3, 2I_3)$ &  51 (27)&  78 (58)&  22 (13)& 100 (100)&   3 (3)&  99 (96)&  25 (4)&  96 (63)&  99 (79)&  98 (69)& 100 (84)\\
$IW_3(5, 3I_3)$  &  36 (29)&  14 (12)&  55 (10)& 100 (100)& 100 (96)&   4 (4)& 100 (100)&  74 (43)&  88 (58)&  86 (54)&  94 (70)\\
$W_3(3, K_3)$ &  93 (44)&  94 (66)&  34 (15)& 100 (100)&  44 (11) & 100 (100)&   3 (3)&  98 (71)& 100 (82)&  99 (74)& 100 (85)\\
$CMT_3(3, K_3)$ &  67 (29)&  42 (20)&  63 (25)& 100 (100)&  89 (42)&  54 (26)&  94 (47)&   3 (4)&   6 (5)&   6 (6)&  14 (8)\\
$CMT_3(5, K_3)$ &  83 (46)&  59 (31)&  81 (41)& 100 (100)&  98 (58)&  73 (37)&  99 (61)&   4 (4)&   3 (3)&   6 (4)&   8 (6)\\
$CMT_3(3, I_3)$ &  78 (38)&  58 (25)&  77 (34)& 100 (100)&  95 (46)&  69 (32)&  97 (51)&   6 (4)&   6 (4)&   3 (4)&   6 (5)\\
$CMT_3(5, I_3)$ &  92 (52)&  79 (39)&  91 (45)& 100 (100)&  99 (65)&  87 (50)& 100 (68)&   9 (6)&   7 (4)&   4 (4)&   3 (4)\\

\hline
\end{tabular}
\end{adjustbox}

\medskip

\begin{adjustbox}{width=\linewidth,center}
\begin{tabular}{l|lllllllllll}
\hline
$n_1=50, n_2=20$ & $W_3(3, I_3)$ & $IW_3(3, I_3)$ & $CMT_3(1, I_3)$ &$CMU_3$ & $W_3(3, 2I_3)$  & $IW_3(5, 3I_3)$ & $W_3(3, K_3)$ & $CMT_3(3, K_3)$ & $CMT_3(5, K_3)$ & $CMT_3(3, I_3)$ & $CMT_3(5, I_3)$ \\ \hline
$W_3(3, I_3)$  &   3 (4) &  26 (30)&  17 (14)& 100 (100)&  28 (5)&  24 (23)&  91 (14)&  95 (76)&  99 (89)&  98 (83)& 100 (92) \\
$IW_3(3, I_3)$  &  15 (11)&   4 (4)&  30 (12)& 100 (100)&  77 (43)&   3 (3)&  97 (58)&  82 (60)&  94 (73)&  92 (67)&  98 (82)\\
$CMT_3(1, I_3)$ &   2  (3) &  11 (4)&   4 (4)& 100 (100)&   3 (3)&  13 (3)&   7 (3)&  91 (64)&  98 (80)&  97 (75)&  99 (86)\\
$CMU_3$ & 100 (100)& 100 (100)& 100 (100)&   5 (5)& 100 (100)& 100 (100)& 100 (100)& 100 (100)& 100 (100)& 100 (100)& 100 (100)\\
$W_3(3, 2I_3)$ &  74 (44)&  91 (79)&  43 (26)& 100 (100)&   3 (4)& 100 (99)&  25 (4)& 100 (87)& 100 (95)& 100 (90)& 100 (97)\\
$IW_3(5, 3I_3)$ &  66 (49)&  27 (20)&  96 (27)& 100 (100)& 100 (99)&   4 (4)& 100 (100)&  93 (70)&  98 (82)&  97 (78)&  99 (90)\\
$W_3(3, K_3)$ &  99 (75)&  99 (85)&  61 (30)& 100 (100)&  68 (21)& 100 (100)&   3 (4)& 100 (88)& 100 (97)& 100 (91)& 100 (97)\\
$CMT_3(3, K_3)$ &  73 (29)&  42 (15)&  72 (27)& 100 (100)&  96 (43)&  56 (22)&  99 (49)&   4 (4)&   7 (6)&  11 (7)&  20 (11)\\
$CMT_3(5, K_3)$ &  93 (49)&  65 (27)&  90 (43)& 100 (100) & 100 (64)&  81 (37)& 100 (68)&   4 (3)&   4 (4)&   8 (5)&  12 (8)\\
$CMT_3(3, I_3)$ &  87 (39)&  64 (21)&  85 (34)& 100 (100)&  99 (53)&  76 (31)& 100 (56)&   6 (4)&   9 (5)&   4 (4)&   7 (7)\\
$CMT_3(5, I_3)$ &  98 (60)&  82 (41)&  96 (59)& 100 (100)& 100 (74)&  94 (46)& 100 (75)&  11 (5)&   7 (4)&   4 (4)&   5 (4)\\

\hline
\end{tabular}
\end{adjustbox}

\medskip 

\begin{adjustbox}{width=\linewidth,center}
\begin{tabular}{l|lllllllllll}
\hline
$n_1=50, n_2=30$ & $W_3(3, I_3)$ & $IW_3(3, I_3)$ & $CMT_3(1, I_3)$ &$CMU_3$ & $W_3(3, 2I_3)$  & $IW_3(5, 3I_3)$ & $W_3(3, K_3)$ & $CMT_3(3, K_3)$ & $CMT_3(5, K_3)$ & $CMT_3(3, I_3)$ & $CMT_3(5, I_3)$  \\ \hline
$W_3(3, I_3)$  &   3 (3)&  34 (34)&  17 (11)& 100 (100)&  62 (20)&  43 (39)& 100 (48)&  98 (81)& 100 (93)&  99 (89)& 100 (96)\\
$IW_3(3, I_3)$  &  27 (23)&   5 (4)&  35 (13)& 100 (100)&  93 (69)&   7 (6)& 100 (83)&  87 (63)&  96 (78)&  96 (71)& 100 (85)\\
$CMT_3(1, I_3)$  &   3 (3)&  21 (8)&   3 (4)& 100 (100)&  10 (4)&  50 (4)&  25 (7)&  96 (70)&  99 (88)&  99 (81)& 100 (92)\\
$CMU_3$  & 100 (100)& 100 (100)& 100 (100)&   5 (5)& 100 (100)& 100 (100)& 100 (100)& 100 (100)& 100 (100)& 100 (100)& 100 (100)\\
$W_3(3, 2I_3)$  &  82 (49)&  95 (82)&  40 (21)& 100 (100)&   3 (3)& 100 (100)&  66 (5)& 100 (92)& 100 (98)& 100 (94)& 100 (99)\\
$IW_3(5, 3I_3)$ &  73 (55)&  25 (17)&  96 (25)& 100 (100)& 100 (100)&   5 (5)& 100 (100)&  95 (74)&  99 (88)&  99 (84)& 100 (93)\\
$W_3(3, K_3)$ & 100 (84)& 100 (91)&  60 (25)& 100 (100)&  86 (20)& 100 (100)&   3 (4)& 100 (93)& 100 (98)& 100 (96)& 100 (100)\\
$CMT_3(3, K_3)$ &  94 (58)&  70 (38)&  91 (52)& 100 (100)& 100 (78)&  85 (48)& 100 (82)&   4 (4)&   7 (6)&  12 (7)&  22 (12)\\
$CMT_3(5, K_3)$  &  99 (81)&  90 (57)&  98 (73)& 100 (100)& 100 (92)&  97 (71)& 100 (95)&   5 (4)&   4 (4)&   9 (5)&  14 (7)\\
$CMT_3(3, I_3)$  &  98 (70)&  87 (46)&  97 (62)& 100 (100)& 100 (84)&  95 (60)& 100 (86)&   9 (4)&  10 (5)&   4 (4)&   8 (6) \\
$CMT_3(5, I_3)$  & 100 (88)&  97 (70)& 100 (83)& 100 (100)& 100 (95)&  99 (81)& 100 (97)&  18 (8)&  12 (5)&   5 (4)&   4 (4)\\

\hline
\end{tabular}
\end{adjustbox}

\medskip 

\begin{adjustbox}{width=\linewidth,center}
\begin{tabular}{l|llllllllllll}
\hline
$n_1=50, n_2=50$ & $W_3(3, I_3)$ & $IW_3(3, I_3)$ & $CMT_3(1, I_3)$ &$CMU_3$ & $W_3(3, 2I_3)$ & $IW_3(5, 3I_3)$ & $W_3(3, K_3)$ & $CMT_3(3, K_3)$ & $CMT_3(5, K_3)$ & $CMT_3(3, I_3)$ & $CMT_3(5, I_3)$\\ \hline
$W_3(3, I_3)$   &   3 (3)&  42 (37)&  11 (6)& 100 (100)&  91 (52)&  76 (67)& 100 (89)& 100 (90) & 100 (97) & 100 (94) & 100 (99)\\
$IW_3(3, I_3)$ &  &   4 (4)&  41 (15)& 100 (100)&  99 (91)&  20 (14)& 100 (97)&  93 (66)&  99 (85)&  99 (79)& 100 (94)\\
$CMT_3(1, I_3)$ &  &  &   3 (4)& 100 (100)&  34 (15)&  98 (18)&  63 (19)&  99 (81)& 100 (94)& 100 (89)& 100 (98)\\
$CMU_3$  &  &  &  &   5 (5)& 100 (100)& 100 (100)& 100 (100)& 100 (100)& 100 (100)& 100 (100)& 100 (100)\\
$W_3(3, 2I_3)$ &  &  &  &  &   3 (3)& 100 (100)&  95 (17)& 100 (97)& 100 (100)& 100 (98)& 100 (100)\\
$IW_3(5, 3I_3)$ &  &  &  &  &  &   5 (5)& 100 (100)&  98 (80)& 100 (93)& 100 (89)& 100 (97)\\
$W_3(3, K_3)$ &  &  &  &  &  &  &   3 (3)& 100 (98)& 100 (100)& 100 (99)& 100 (100)\\ 
$CMT_3(3, K_3)$  &  &  &  &  &  &  &  &   4 (4)&   8 (6)&  15 (6)&  32 (14)\\
$CMT_3(5, K_3)$  &  &  &  &  &  &  &  &  &   4 (4)&  13 (6)&  20 (7)\\
$CMT_3(3, I_3)$  &  &  &  &  &  &  &  &  &  &   4 (3)&   8 (6)\\
$CMT_3(5, I_3)$ &  &  &  &  &  &   &   &  &  &  & 4 (3)\\

\hline
\end{tabular}
\end{adjustbox}
\end{table}
\newpage

\begin{table}[htbp] 
\caption{{ The percentage of rejected $H_0$} for different sample sizes for $2\times 2$ matrices, $\nu=5, \omega = I_d (\nu=5, \omega=2I_d)$.}\label{pow2eye5}
\begin{adjustbox}{width=\linewidth,center}
\begin{tabular}{l|lllllllllll}
\hline
$n_1=20, n_2=20$ & $W_{2}(2.5, I_2)$ & $IW_{2}(2.5, I_2)$ & $CMT_2(1, I_2)$ & $CMU_2$ & $W_{2}(2.5, 2I_2)$ & $IW_{2}(4, 2.5I_2)$ & $W_2(2.5, K_2)$ & $CMT_2(3, K_2)$ & $CMT_2(5, K_2)$ & $CMT_2(3, I_2)$ & $CMT_2(5, I_2)$ \\ \hline
$W_{2}(2.5, I_2)$ & 5 (6) & 16 (12) & 12 (13) & 100 (100) & 10 (8)& 6 (5) & 17 (12) & 36 (28) & 41 (33) & 56 (41) & 60 (50) \\
$IW_{2}(2.5, I_2)$ &  & 5 (4) & 11 (11) & 100 (100) & 45 (31) & 14 (14)& 8 (8) & 30 (23)& 37 (28)& 49 (40)& 55 (47) \\
$CMT_2(1, I_2)$ &  &  & 4 (4)& 100 (99) & 16 (15) & 14 (14)& 7 (8)& 13 (10)& 17 (13)& 25 (22)& 36 (28)\\
$CMU_2$ &  &  &  & 5 (5)& 100 (100)& 100 (100)& 100 (100)& 96 (90)& 94 (86)& 87 (74) & 82 (66) \\
$W_{2}(2.5, 2I_2)$ &  &  &  &  & 6 (6)& 49 (38)& 29 (20) & 39 (31)& 47 (35)& 58 (48)& 66 (52) \\
$IW_{2}(4, 2.5I_2)$ &  &  &  &  &  & 5 (5)& 16 (14)& 35 (28)& 43 (33)& 53 (44)& 60 (50)\\
$W_2(2.5, K_2)$ &  &  &  &  &  &  & 5 (5)& 26 (20)& 30 (24)& 42 (35)& 51 (41)\\
$CMT_2(3, K_2)$ &  &  &  &  &  &  &  & 5 (4) & 5 (5)& 8 (8)& 12 (11)\\
$CMT_2(5, K_2)$ &  &  &  &  &  &  &  &  & 5 (4)& 6 (7)& 9 (8)\\
$CMT_2(3, I_2)$ &  &  &  &  &  &  &  &  &  & 5 (5)& 5 (5)\\
$CMT_2(5, I_2)$ &  &  &  &  &  &  &  &  &  &  & 5 (4)\\ \hline
\end{tabular}
\end{adjustbox}

\medskip 
\begin{adjustbox}{width=\linewidth,center}
\begin{tabular}{l|lllllllllll}
\hline
$n_1=30, n_2=20$ & $W_{2}(2.5, I_2)$ & $IW_{2}(2.5, I_2)$ & $CMT_2(1, I_2)$ & $CMU_2$ & $W_{2}(2.5, 2I_2)$ & $IW_{2}(4, 2.5I_2)$ & $W_2(2.5, K_2)$ & $CMT_2(3, K_2)$ & $CMT_2(5, K_2)$ & $CMT_2(3, I_2)$ & $CMT_2(5, I_2)$ \\ \hline
$W_{2}(2.5, I_2)$ & 5 (6) & 19 (14)& 21 (18)& 100 (100)& 8 (6) & 5 (4) & 24 (19)& 52 (41)& 58 (47)& 69 (60)& 76 (65)\\
$IW_{2}(2.5, I_2)$ & 18 (12)& 4 (4)& 16 (17)& 100 (100)& 49 (33)& 12 (11)& 12 (12)& 45 (40)& 50 (45)& 64 (54)& 70 (63)\\
$CMT_2(1, I_2)$ & 10 (9)& 7 (8)& 4 (4)& 100 (100)& 14 (11)& 11 (9)& 5 (6)& 19 (15)& 23 (18)& 37 (30)& 45 (37)\\
$CMU_2$ & 100 (100)& 100 (100)& 100 (100)& 5 (5)& 100 (100)& 100 (100)& 100 (100)& 99 (93)& 97 (92)& 92 (82)& 90 (78)\\
$W_{2}(2.5, 2I_2)$ & 15 (12)& 57 (47)& 25 (21)& 100 (100)& 6 (6)& 62 (46)& 40 (31)& 57 (47)& 64 (49)& 73 (62)& 80 (70)\\
$IW_{2}(4, 2.5I_2)$ & 10 (7)& 20 (21)& 23 (20)& 100 (100)& 60 (45)& 4 (5)& 26 (23)& 52 (41)& 59 (47)& 70 (60)& 77 (67)\\
$W_2(2.5, K_2)$ & 14 (10)& 5 (5)& 11 (12)& 100 (100)& 25 (15)& 13 (10)& 4 (4)& 37 (32)& 43 (36) & 56 (51)& 64 (56)\\
$CMT_2(3, K_2)$ & 35 (29)& 30 (24)& 13 (10) & 99 (95)& 44 (31)& 37 (29)& 23 (20)& 4 (5)& 5 (5)& 11 (11)& 15 (13)\\
$CMT_2(5, K_2)$ & 45 (34)& 35 (27)& 17 (13)& 98 (94)& 53 (37)& 44 (33)& 30 (24)& 5 (4)& 5 (5)& 9 (8)& 12 (12)\\
$CMT_2(3, I_2)$ & 59 (48) & 51 (40)& 31 (24)& 94 (83)& 65 (46)& 60 (45)& 44 (36)& 8 (7)& 7 (5)& 4 (5)& 6 (5)\\
$CMT_2(5, I_2)$ & 67 (54)& 61 (50)& 36 (28)& 89 (78)& 73 (57)& 69 (54)& 55 (45)& 11 (10)& 9 (8)& 5 (5)& 5 (5)\\ \hline
\end{tabular}
\end{adjustbox} 

\medskip

\begin{adjustbox}{width=\linewidth,center}
\begin{tabular}{l|lllllllllll}
\hline
$n_1=50, n_2=20$ & $W_{2}(2.5, I_2)$ & $IW_{2}(2.5, I_2)$ & $CMT_2(1, I_2)$ & $CMU_2$ & $W_{2}(2.5, 2I_2)$ & $IW_{2}(4, 2.5I_2)$ & $W_2(2.5, K_2)$ & $CMT_2(3, K_2)$ & $CMT_2(5, K_2)$ & $CMT_2(3, I_2)$ & $CMT_2(5, I_2)$ \\ \hline
$W_{2}(2.5, I_2)$ & 5 (5)& 26 (17)& 32 (29)& 100 (100)& 5 (4)& 4 (3)& 37 (27)& 69 (60)& 77 (65)& 85 (77)& 89 (81)\\
$IW_{2}(2.5, I_2)$ & 19 (13)& 4 (5)& 25 (26)& 100 (100)& 52 (37)& 8 (9)& 17 (18)& 61 (54)& 69 (62)& 80 (73)& 86 (78)\\
$CMT_2(1, I_2)$ & 8 (5)& 5 (4)& 4 (4)& 100 (100)& 10 (16)& 6 (7)& 4 (5)& 23 (20)& 31 (24)& 47 (41)& 59 (50)\\
$CMU_2$ & 100 (100) & 100 (100)& 100 (100) & 5 (5) & 100 (100) & 100 (100) & 100 (100) & 100 (100) & 99 (96)& 97 (89) & 95 (84) \\
$W_{2}(2.5, 2I_2)$ & 20 (15)& 73 (62) & 38 (33) & 100 (100) & 5 (6) & 73 (51) & 65 (48) & 76 (65) & 83 (70) & 88 (79) & 92 (83)\\
$IW_{2}(4, 2.5I_2)$ & 13 (11) & 29 (31)& 34 (31)& 100 (100)& 72 (57)& 5 (5)& 43 (35)& 72 (60)& 75 (66)& 85 (77)& 89 (82)\\
$W_2(2.5, K_2)$ & 13 (8) & 4 (5)& 17 (18)& 100 (100) & 26 (14)& 10 (7)& 4 (5) & 51 (46) & 58 (52)& 71 (66)& 79 (72)\\
$CMT_2(3, K_2)$ & 37 (28) & 29 (22)& 12 (8)& 100 (99) & 46 (30) & 39 (27)& 23 (18) & 5 (4)& 5 (6)& 14 (14)& 19 (18)\\
$CMT_2(5, K_2)$ & 47 (32) & 36 (28)& 17 (12)& 100 (98)& 56 (36) & 48 (33)& 31 (23)& 4 (4)& 4 (5)& 11 (10) & 16 (15)\\
$CMT_2(3, I_2)$ & 68 (48)& 57 (46)& 34 (23)& 99 (92)& 72 (57)& 66 (53) & 49 (38)& 8 (8)& 6 (5)& 5 (5)& 6 (6)\\
$CMT_2(5, I_2)$ & 79 (60)& 70 (57)& 43 (32)& 97 (87)& 84 (64)& 79 (62)& 59 (47)& 11 (11)& 9 (7)& 5 (4)& 5 (5)\\ \hline
\end{tabular}\end{adjustbox}

\medskip 

\begin{adjustbox}{width=\linewidth,center}
\begin{tabular}{l|lllllllllll}
\hline
$n_1=50, n_2=30$ & $W_{2}(2.5, I_2)$ & $IW_{2}(2.5, I_2)$ & $CMT_2(1, I_2)$ & $CMU_2$ & $W_{2}(2.5, 2I_2)$ & $IW_{2}(4, 2.5I_2)$ & $W_2(2.5, K_2)$ & $CMT_2(3, K_2)$ & $CMT_2(5, K_2)$ & $CMT_2(3, I_2)$ & $CMT_2(5, I_2)$ \\ \hline
$W_{2}(2.5, I_2)$ & 5 (5) & 30 (21) & 32 (29)& 100 (100) & 10 (7) & 5 (4)& 44 (29)& 76 (64)& 82 (72)& 91 (83) & 95 (89)\\
$IW_{2}(2.5, I_2)$ & 25 (19)& 5 (5) & 23 (23) & 100 (100) & 73 (60)& 18 (16)& 19 (17)& 65 (58)& 74 (65)& 85 (79) & 92 (86) \\
$CMT_2(1, I_2)$ & 16 (12)& 9 (9)& 4 (4)& 100 (100)& 23 (16)& 15 (14)& 6 (5)& 29 (21)& 34 (27)& 58 (46)& 67 (58)\\
$CMU_2$ & 100 (100)& 100 (100)& 100 (100)& 5 (5)& 100 (100)& 100 (100)& 100 (100)& 100 (100)& 100 (99)& 100 (97)& 99 (95)\\
$W_{2}(2.5, 2I_2)$ & 20 (14)& 82 (71) & 39 (32) & 100 (100)& 6 (6) & 85 (69) & 73 (49) & 81 (71) & 86 (74) & 93 (85) & 97 (90) \\
$IW_{2}(4, 2.5I_2)$ & 14 (8) & 29 (29) & 34 (29) & 100 (100) & 87 (75) & 5 (5) & 44 (35) & 76 (64) & 83 (73) & 92 (84) & 95 (89) \\
$W_2(2.5, K_2)$ & 28 (17) & 8 (7) & 14 (16) & 100 (100) & 49 (28) & 22 (17) & 4 (4) & 54 (49) & 63 (57)& 77 (71) & 85 (79)\\
$CMT_2(3, K_2)$ & 63 (46) & 51 (40) & 20 (17) & 100 (100) & 68 (53) & 62 (45) & 42 (30) & 5  (4) & 6 (5) & 16 (14) & 23 (19) \\
$CMT_2(5, K_2)$ & 72 (58) & 63 (49) & 28 (20) & 100  (100) & 80 (62) & 73 (58) & 50 (38) & 5 (4) & 5 (4) & 12 (10) & 17 (16) \\
$CMT_2(3, I_2)$ & 86 (74) & 80 (68) & 46 (37) & 100 (98) & 90 (77)& 86 (75) & 70 (60)& 10 (10)& 9 (8) & 5 (5) & 6 (5) \\
$CMT_2(5, I_2)$ & 93 (81) & 87 (78) & 60 (47)& 100 (96) & 95 (85) & 93 (83) & 79 (68) & 17 (14) & 14 (11) & 5 (5)& 5 (5) \\ \hline
\end{tabular}\end{adjustbox}

\medskip 

\begin{adjustbox}{width=\linewidth,center}
\begin{tabular}{l|lllllllllll}
\hline
$n_1=50, n_2=50$ & $W_{2}(2.5, I_2)$ & $IW_{2}(2.5, I_2)$ & $CMT_2(1, I_2)$ & $CMU_2$ & $W_{2}(2.5, 2I_2)$ & $IW_{2}(4, 2.5I_2)$ & $W_2(2.5, K_2)$ & $CMT_2(3, K_2)$ & $CMT_2(5, K_2)$ & $CMT_2(3, I_2)$ & $CMT_2(5, I_2)$ \\ \hline
$W_{2}(2.5, I_2)$ & 5 (5) & 41 (29)& 30 (25) & 100 (100) & 19 (12)& 11 (7)& 48 (28)& 85 (74)& 91 (79)& 97 (91) & 99 (95)\\
$IW_{2}(2.5, I_2)$ &  & 4 (5)& 21 (20) & 100 (10)& 91 (82)& 30 (32)& 16 (14)& 74 (65)& 84 (74)& 93 (89)& 96 (93)\\
$CMT_2(1, I_2)$ &  &  & 5 (4)& 100 (100)& 43 (29)& 32 (27)& 14 (13)& 33 (25)& 41 (30)& 65 (54)& 79 (67)\\
$CMU_2$ &  &  &  & 5 (5)& 100 (100)& 100 (100)& 100 (100)& 100 (100)& 100 (100)& 100 (100)& 100 (99)\\
$W_{2}(2.5, 2I_2)$ &  &  &  &  & 5 (6)& 96 (89)& 80 (54)& 89 (77)& 94 (85)& 98 (92)& 99 (96)\\
$IW_{2}(4, 2.5I_2)$ &  &  &  &  &  & 5 (5)& 46 (32)& 84 (72)& 90 (82)& 97 (93)& 99 (95)\\
$W_2(2.5, K_2)$ &  &  &  &  &  &  & 5 (4)& 60 (52)& 74 (64)& 87 (80)& 92 (86)\\
$CMT_2(3, K_2)$ &  &  &  &  &  &  &  & 5 (5)& 5 (5)& 17 (15)& 24 (22)\\
$CMT_2(5, K_2)$ &  &  &  &  &  &  &  &  & 5 (5)& 11 (11)& 18 (17)\\
$CMT_2(3, I_2)$ &  &  &  &  &  &  &  &  &  & 5 (4)& 6 (5)\\
$CMT_2(5, I_2)$ &  &  &  &  &  &  &  &  &  &  & 5 (4)\\ \hline
\end{tabular}
\end{adjustbox}
\end{table}
\newpage

\begin{table}[htbp]
\caption{{ The percentage of rejected $H_0$} for different sample sizes for $3\times 3$ matrices,  $\nu=5, \omega = I_d (\nu=5, \omega=2I_d)$.}\label{pow3eye52}
\begin{adjustbox}{width=\linewidth,center}
\begin{tabular}{l|lllllllllll}
\hline
$n_1=20, n_2=20$& $W_3(3, I_3)$& $IW_3(3, I_3)$& $CMT_3(1, I_3)$& $CMU_3$& $W_3(3, 2I_3)$& $IW_3(5, 3I_3)$& $W_3(3, K_3)$& $CMT_3(3, K_3)$& $CMT_3(5, K_3)$& $CMT_3(3, I_3)$& $CMT_3(5, I_3)$ \\ \hline
$W_3(3, I_3)$& 5 (6)& 11 (9)& 6 (8)& 100 (100)& 7 (7)& 11 (8)& 8 (7)& 13 (10)& 14 (12)& 14 (9)& 15 (12)\\
$IW_3(3, I_3)$& & 4 (5)& 6 (6)& 100 (100)& 15 (12)& 7 (8)& 14 (12)& 10 (9)& 11 (10)& 12 (9)& 13 (11)\\
$CMT_3(1, I_3)$& & & 6 (6)& 100 (100)& 7 (7)& 7 (6)& 8 (7)& 10 (9)& 13 (10)& 12 (9)& 13 (10)\\
$CMU_3$& & & & 6 (5)& 100 (100)& 100 (100)& 100 (100)& 100 (100)& 100 (100)& 100 (100)& 100 (100)\\
$W_3(3, 2I_3)$& & & & & 6 (6)& 36 (25)& 6 (7)& 13 (10)& 16 (11)& 14 (11)& 17 (12)\\
$IW_3(5, 3I_3)$& & & & & & 4 (4)& 41 (28)& 12 (9)& 13 (10)& 13 (10)& 15 (11)\\
$W_3(3, K_3)$& & & & & & & 6 (6)& 13 (10)& 16 (10)& 14 (10)& 17 (12)\\
$CMT_3(3, K_3)$& & & & & & & & 5 (5)& 5 (5)& 4 (5)& 5 (5)\\
$CMT_3(5, K_3)$& & & & & & & & & 5 (5)& 5 (5)& 5 (5)\\
$CMT_3(3, I_3)$& & & & & & & & & & 4 (5)& 5 (5)\\
$CMT_3(5, I_3)$& & & & & & & & & & & 4 (5)\\ \hline
\end{tabular}

\end{adjustbox}

\medskip 

\begin{adjustbox}{width=\linewidth,center}
\begin{tabular}{l|lllllllllll}
\hline
$n_1=30, n_2=20$& $W_3(3, I_3)$& $IW_3(3, I_3)$& $CMT_3(1, I_3)$& $CMU_3$& $W_3(3, 2I_3)$& $IW_3(5, 3I_3)$& $W_3(3, K_3)$& $CMT_3(3, K_3)$& $CMT_3(5, K_3)$& $CMT_3(3, I_3)$& $CMT_3(5, I_3)$ \\ \hline
$W_3(3, I_3)$& 5 (6)& 16 (13)& 9 (8)&100 (100)&5 (4)&10 (5)&5 (5)&20 (16)&24 (18)&22 (16)&24 (18) \\
$IW_3(3, I_3)$&8 (6)&4 (5)&6 (7)&100 (100)&12 (8)&5 (5)&12 (9)&16 (14)&21 (16)&19 (14)&22 (17) \\
$CMT_3(1, I_3)$&5 (5)&5 (4)&6 (6)&100 (100)&5 (5)&4 (4)&5 (5)&17 (13)&20 (14)&17 (14)&21 (16) \\
$CMU_3$&100 (100)&100 (100)&100 (100)& 5 (5)&100 (100)&100 (100)&100 (100)&100 (100)&100 (100)&100 (100)&100 (100) \\
$W_3(3, 2I_3)$&11 (10)&23 (18)&11 (10)&100 (100)&6 (6)&51 (36)&5 (5)&21 (16)&25 (18)&22 (15)&28 (21) \\
$IW_3(5, 3I_3)$&13 (9)&12 (13)&9 (9)&100 (100)&37 (24)&5 (5)&44 (26)&20 (15)&22 (17)&20 (16)&25 (17) \\
$W_3(3, K_3)$&12 (11)&24 (18)&11 (10)&100 (100)&8 (8)&59 (46)&6 (6)&21 (16)&26 (17)&23 (17)&25 (18) \\
$CMT_3(3, K_3)$&10 (6)&7 (7)&8 (6)&100 (100)&12 (7)&10 (7)&11 (8)&4 (5)&5 (5)&5 (6)&6 (7) \\
$CMT_3(5, K_3)$&12 (7)&10 (7)&9 (7)&100 (100)&12 (9)&10 (8)&11 (8)&4 (5)&5 (5)&6 (5)&5 (6) \\
$CMT_3(3, I_3)$&9 (6)&8 (6)&9 (7)&100 (100)&10 (7)&10 (7)&11 (8)&5 (5)&4 (4)&5 (5)&5 (6) \\
$CMT_3(5, I_3)$&14 (8)&10 (7)&11 (7)&100 (100)&13 (9)&12 (9)&13 (8)&5 (5)&4 (5)&4 (5)&5 (5) \\ \hline
\end{tabular}
\end{adjustbox}

\medskip

\begin{adjustbox}{width=\linewidth,center}
\begin{tabular}{l|lllllllllll}
\hline
$n_1=50, n_2=20$ & $W_3(3, I_3)$ & $IW_3(3, I_3)$ & $CMT_3(1, I_3)$ & $CMU_3$ & $W_3(3, 2I_3)$ & $IW_3(5, 3I_3)$ & $W_3(3, K_3)$ & $CMT_3(3, K_3)$ & $CMT_3(5, K_3)$ & $CMT_3(3, I_3)$ & $CMT_3(5, I_3)$ \\ \hline
$W_3(3, I_3)$ &5 (5) &25 (19) &13 (12) &100 (100) &4 (4) &12 (5) & 4 (4) &32 (25) &40 (30) &35 (27) &42 (31) \\
$IW_3(3, I_3)$ &5 (4) &4 (4) &7 (8) &100 (100) &8 (5) &3 (3) &8 (5) &26 (23) &34 (26) &30 (25) &36 (27) \\
$CMT_3(1, I_3)$ &3 (3) &4 (3) &5 (5) &100 (100) &3 (3) &3 (3) &3 (3) &26 (18) &32 (22) &29 (22) &36 (25) \\
$CMU_3$ &100 (100) &100 (100) &100 (100) &5 (5) &100 (100) &100 (100) &100 (100) &100 (100) &100 (100) &100 (100) &100 (100) \\
$W_3(3, 2I_3)$ & 15 (14) &36 (31) &15 (14) &100 (100) &5 (5) &71 (55) &3 (3) &35 (26) &42 (29) &39 (28) &46 (31) \\
$IW_3(5, 3I_3)$ &13 (9) &20 (20) &13 (13) &100 (100) &40 (26) &5 (4) &43 (28) &32 (25) &40 (28) &33 (27) &42 (31) \\
$W_3(3, K_3)$ &18 (18) &38 (31) &16 (13) &100 (100) &10 (11) &78 (66) &5 (5) &35 (26) & 43 (28) &38 (27) &43 (31) \\
$CMT_3(3, K_3)$ &6 (4) &4 (3) &5 (4) &100 (100) &6 (4) &5 (4) &7 (4) &5 (5) &5 (6) &6 (7) &7 (7) \\
$CMT_3(5, K_3)$ &7 (4) &6 (3) & 7 (4) &100 (100) &7 (5) &6 (5) &7 (4) &4 (4) &4 (5) &4 (6) &5 (6) \\
$CMT_3(3, I_3)$ &6 (4) &5 (4) &6 (4) &100 (100) &7 (4) &6 (5) & 7 (5) &4 (4) &4 (4) &4 (4) &5 (5) \\
$CMT_3(5, I_3)$ &9 (5) &7 (4) &7 (5) &100 (100) &9 (5) &8 (4) &10 (6) &4 (3) &4 (3) &4 (4) &4 (5) \\ \hline
\end{tabular}
\end{adjustbox}

\medskip

\begin{adjustbox}{width=\linewidth,center}
\begin{tabular}{l|lllllllllll}
\hline
$n_1=50, n_2=30$ & $W_3(3, I_3)$ & $IW_3(3, I_3)$ & $CMT_3(1, I_3)$ & $CMU_3$ & $W_3(3, 2I_3)$ & $IW_3(5, 3I_3)$ & $W_3(3, K_3)$ & $CMT_3(3, K_3)$ & $CMT_3(5, K_3)$ & $CMT_3(3, I_3)$ & $CMT_3(5, I_3)$ \\ \hline
$W_3(3, I_3)$ &5 (5) &24 (18) &11 (10) &100 (100) &5 (4) &15 (6) &5 (5) &32 (21) &39 (26) &34 (23) &41 (29) \\
$IW_3(3, I_3)$ &11 (9) &4 (5) &7 (8) &100 (100) &19 (12) &6 (5) &18 (10) &25 (19) &30 (22) &28 (20) &34 (26) \\
$CMT_3(1, I_3)$ &4 (4) &4 (4) &5 (6) &100 (100) &4 (5) &4 (5) &5 (5) &24 (17) &32 (20) &28 (20) &35 (25) \\
$CMU_3$ &100 (100) &100 (100) &100 (100) &5 (5) &100 (100) &100 (100) &100 (100) &100 (100) &100 (100) &100 (100) &100 (100) \\
$W_3(3, 2I_3)$ &14 (11) &38 (26) &12 (10) &100 (100) &5 (6) &80 (63) &5 (5) &33 (22) &38 (27) &37 (24) &45 (27) \\
$IW_3(5, 3I_3)$ &19 (10) &18 (16) &10 (10) &100 (100) &67 (44) &4 (4) &72 (53) &30 (22) &37 (26) &31 (22) &40 (27) \\
$W_3(3, K_3)$ &16 (11) &36 (29) &13 (11) &100 (100) &9 (8) &85 (74) &6 (6) &32 (22) &41 (25) &35 (24) &44 (26) \\
$CMT_3(3, K_3)$ &12 (8) &9 (6) &11 (7) &100 (100) &15 (8) &12 (8) &14 (8) &4 (5) &5 (5) &6 (6) &7 (7) \\
$CMT_3(5, K_3)$ &16 (9) &13 (9) &14 (9) &100 (100) &19 (10) &14 (9) &17 (9) &4 (4) &4 (5) &5 (5) &7 (6) \\
$CMT_3(3, I_3)$ &15 (8) &10 (8) &12 (8) &100 (100) &15 (8) &15 (9) &15 (9) &4 (4) &4 (4) &4 (5) &6 (5) \\
$CMT_3(5, I_3)$ &21 (11) &16 (10) &16 (10) &100 (100) &22 (11) &20 (13) &22 (12) &4 (4) &4 (4) &4 (4) &4 (5) \\ \hline
\end{tabular}
\end{adjustbox}

\medskip 

\begin{adjustbox}{width=\linewidth,center}
\begin{tabular}{l|lllllllllll}
\hline
$n_1=50, n_2=50$ & $W_3(3, I_3)$ & $IW_3(3, I_3)$ & $CMT_3(1, I_3)$ & $CMU_3$ & $W_3(3, 2I_3)$ & $IW_3(5, 3I_3)$ & $W_3(3, K_3)$ & $CMT_3(3, K_3)$ & $CMT_3(5, K_3)$ & $CMT_3(3, I_3)$ & $CMT_3(5, I_3)$ \\ \hline
$W_3(3, I_3)$ & 5 (5)& 22 (17) & 7 (7)& 100 (100)& 8 (7)& 24 (11)& 10 (7)& 28 (18)& 38 (21)& 33 (18)& 42 (25)\\
$IW_3(3, I_3)$ &   & 4 (4)& 6 (6)& 100 (100)& 37 (26)& 12 (13)& 39 (27)& 21 (15)& 26 (19)& 25 (17)& 33 (20)\\
$CMT_3(1, I_3)$ &   &   & 5 (6)& 100 (100)& 9 (8)& 6 (7)& 9 (8)& 23 (14)& 30 (18)& 27 (17)& 35 (21)\\
$CMU_3$ &   &   &   & 5 (5)& 100 (100)& 100 (100)& 100 (100)& 100 (100)& 100 (100)& 100 (100)& 100 (100)\\
$W_3(3, 2I_3)$ &   &   &   &   & 6 (5)& 89 (72)& 7 (6)& 31 (17)& 45 (23)& 35 (22)& 48 (26)\\
$IW_3(5, 3I_3)$ &   &   &   &   &   & 4 (5)& 95 (83)& 28 (17)& 38 (21)& 30 (18)& 43 (22)\\
$W_3(3, K_3)$ &   &   &   &   &   &   & 6 (6)& 30 (16)& 40 (22)& 35 (20)& 48 (27)\\
$CMT_3(3, K_3)$ &   &   &   &   &   &   &   & 5 (4)& 5 (4)& 5 (5)& 7 (6)\\
$CMT_3(5, K_3)$ &   &   &   &   &   &   &   &   & 4 (5)& 4 (4)& 5 (5)\\
$CMT_3(3, I_3)$ &   &   &   &   &   &   &   &   &   & 4 (4)& 5 (4)\\
$CMT_3(5, I_3)$ &   &   &   &   &   &   &   &   &   &   & 4 (4)\\ \hline
\end{tabular}
\end{adjustbox}
\end{table}
\newpage

\section{Real data examples}\label{sec::realdata} 

\subsection{Cryptocurrency data example}
Recently, the one-minute cryptocurrency data connected to important Bitcoin events  \cite{fruehwirt2021cumulation} has been analyzed in \cite{nas2, nas1}. 
 
Following the methodology outlined in \cite{nas1}, we selected two-day periods and computed the minute logarithmic returns. We computed covariance matrices for every hour, resulting in a total of $n = 48$ covariance matrices, with 24 for the first day and 24 for the second day. Since the test for $\nu = 1$ and $\Sigma = I_2$ is the most powerful, only the results for these parameters are given for the sake of brevity.
 
The results presented in Table \ref{pvals1min} indicate that our test has detected a statistically significant difference in the distribution of covariance matrices on the day before and the day of the occurrence of the event in most cases. The test consistently detects the statistically significant difference without any failures. These findings align with those outlined in \cite{nas1} and \cite{nas2}, further highlighting the potential financial application.

\begin{table}[htbp]
\caption{$p$-values of testing the change in covariance structure before and after the Bitcoin important events - 1 minute data}
\label{pvals1min}
\centering
\begin{adjustbox}{width=\linewidth,center}
\begin{tabular}{@{}lp{2.5in}lll}
\hline
Date of event ($T_0$) & Event description & $p_{[T_0-2D, T_0-1D]}$ & $p_{[T_0-1D, T_0]}$ & $p_{[T_0,  T_0+1D]}$  \\ \hline
8 November 2017 & Developers cancel splitting of Bitcoin. &0.04  & 0.09 & 0.20  \\
28 December 2017 & South Korea announces strong measures to regulate the trading of cryptocurrencies. & 0.48 & 0.02 & 0.09 \\
13 January 2018 & Announcement that 80\% of Bitcoin has been mined. & 0.04 & 0.01 & 0.06 \\
30 January 2018 & Facebook bans advertisements promoting cryptocurrencies. & 0.73 & 0.03 & 0.79  \\
7 March 2018 & The US Securities and Exchange Commission says it is necessary for crypto trading platforms to register. & 0 & 0.02 & 0.61 \\
14 March 2018 & Google bans advertisements promoting cryptocurrencies. & 0.43 & 0.24 & 0.03  \\ \hline 
\end{tabular}
\end{adjustbox}
\end{table}


 
\subsection{Non-life insurance data}
The data set 'Insurance' from the R package \textbf{splm} \cite{millo2012splm} was considered for the first time in \cite{millo2011non}. The data consists of the insurance consumption data across Italian provinces during a 5-year period from 1998 to 2001. Following the approach in \cite{tomarchio2021matrix}, we have computed the empirical covariance matrices per province of the following covariates:

\begin{enumerate}
    \item Real per-capita non-life premiums in 2000 Euros (PPCD);
    \item Density of insurance agencies per 1000 inhabitants (AGEN);
    \item Real per-capita GDP (RGDP);
\end{enumerate}

We have split the data into two groups based on geographic location. The first sample consists of $n_1 = 67$ empirical $3\times 3$ covariance matrices corresponding to the Northern provinces, while the second sample consists of $n_2=36$ empirical $3\times 3$ covariance matrices corresponding to the Southern and Island provinces. The results in \cite{tomarchio2021matrix}  corroborate the findings of \cite{millo2011non}. It is suggested that there exists a clear separation between Central-Northern Italy and Southern-Insular Italy data. Therefore, we have tested for the difference in distribution of the covariance matrices between Northern and Southern-Insular data. The p-values of the tests are presented in Table \ref{insurance}. Since every test reports a $p$-value lower than $\alpha = 0.05$, we reject the null hypothesis that the covariance matrices of the selected covariates for Northern and Southern-Insular data are equally distributed, providing further evidence that the regional differences explored in \cite{millo2011non} and \cite{tomarchio2021matrix} exist.

\begin{table}[htbp]
\caption{$p$-values of the novel tests for Italy insurance data}
\centering
\label{insurance}
\begin{tabular}{@{}lllllll@{}}
\toprule
Parameters & $\nu=1, \Sigma = I_5$ & $\nu=1, \Sigma = 2I_5$ & $\nu=2, \Sigma = I_5$ & $\nu=2, \Sigma = 2I_5$ & $\nu=5, \Sigma = I_5$ & $\nu = 5, \Sigma = 2I_5$ \\ \midrule
$p$-value & 0 & 0 & 0 & 0 & 0.0025 & 0.0020 \\ \bottomrule
\end{tabular}
\end{table}

\section*{Conclusion}
In this study, we have developed a novel test for determining equality in distribution of { symmetric positive definite} matrix distributions. Our approach utilizes the { test statistic, which is constructed as the integral of the} squared difference of the empirical Laplace transforms with respect to the noncentral Wishart measure. The Laplace transform-based test is the first of its kind. Through an extensive power study, we have assessed the performance of the test and determined the optimal choice of parameters.
{ The test shows moderate to high powers for the optimal choice of parameters. Our test exhibits the expected behavior in a well-known theoretical scenario.}

Additionally, we have demonstrated the applicability of the test on financial and non-life insurance data, showcasing its effectiveness in practical scenarios.

\section*{Declaration of interest}
The author has no interest to declare.

\section*{Acknowledgements}
The author would like to express his deepest gratitude to Professor Bojana Milošević for providing the useful comments which have significantly improved the outline of the paper. { The author extends his gratitude to the anonymous reviewer, whose comments significantly improved the quality of this paper.}

\providecommand{\bysame}{\leavevmode\hbox to3em{\hrulefill}\thinspace}
\providecommand{\MR}{\relax\ifhmode\unskip\space\fi MR }
\providecommand{\MRhref}[2]{%
  \href{http://www.ams.org/mathscinet-getitem?mr=#1}{#2}
}
\providecommand{\href}[2]{#2}

{\section*{Appendix I}\label{sec::appendix}

In this section, we present the 95-th empirical percentiles of the empirical distributions of the appropriately scaled test statistic $\frac{n_1n_2}{n_1+n_2}L_{n_1, n_2, \nu, \Sigma, \omega}$. This is done to illustrate the method for empirically assessing the asymptotic properties of the novel test. The $N=1000$ values of the statistic $\frac{n_1n_2}{n_1+n_2}L_{n_1, n_2, \nu, \Sigma, \omega}$ have been obtained. The constant $c_{n_1, n_2}=\frac{n_1+n_2}{n_1n_2}$ has been used in assessing the asymptotic behavior of a similar type test \cite{nas1}. We aim to tabulate the values if the null hypothesis holds. In the case of dimension 2, we assume that both samples come from the $W_2(2.5, I_2)$ Wishart distribution, while in the case of dimension 3, we assume that both samples come from the $W_3(3, I_3)$ Wishart distribution. As in Section \ref{sec::power}, we fix the parameter $\Sigma = I_d$ for simplicity.

\begin{table}[ht]
\caption{Empirical 95-th percentiles of the distribution of the scaled statistics. The case of $2\times 2$ matrices.}
\label{asym2}
\centering
\begin{tabular}{llllll}
\hline
 & $n_1=n_2=100$ & $n_1=n_2=200$ & $n_1=n_2=500$ & $n_1=n_2=750$ & $n_1=n_2=1000$ \\ \hline
$\nu=1, \; \omega = I_2$ & 0.0495 & 0.0524 & 0.0549 & 0.0497 & 0.0518 \\
$\nu=1, \; \omega = 2I_2$ & 0.0196 & 0.0208 & 0.0218 & 0.0196 & 0.0212 \\
$\nu=2, \; \omega = I_2$ & 0.0203 & 0.0216 & 0.0223 & 0.0203 & 0.0219 \\
$\nu=2, \; \omega = 2I_2$ & 0.0084 & 0.0090 & 0.0089 & 0.0082 & 0.0086 \\
$\nu=5, \; \omega = I_2$ & 0.0030 & 0.0028 & 0.0029 & 0.0027 & 0.0027 \\
$\nu=5, \; \omega = 2I_2$ & 0.0017 & 0.0014 & 0.0014 & 0.0014 & 0.0013\\ \hline
\end{tabular}
\end{table}

\begin{table}[htbp]
\caption{Empirical 95-th percentiles of the distribution of the scaled statistics. The case of $3\times 3$ matrices.}
\label{asym3}
\begin{tabular}{llllll}
\hline
 & $n_1=n_2=100$ & $n_1=n_2=200$ & $n_1=n_2=500$ & $n_1=n_2=750$ & $n_1=n_2=1000$ \\ \hline
$\nu=1, \; \omega = I_3$ & 0.0109 & 0.0108 & 0.0115 & 0.0107 & 0.0108 \\
$\nu=1, \; \omega = 2I_3$ & 0.0021 & 0.0021 & 0.0022 & 0.0020 & 0.0020 \\
$\nu=2, \; \omega = I_3$ & 0.0016 & 0.0016 & 0.0017 & 0.0015 & 0.0016 \\
$\nu=2, \; \omega = 2I_3$ & 3.410$\times 10^{-4}$ & 3.290$\times 10^{-4}$ & 3.404$\times 10^{-4}$ & 3.122$\times 10^{-4}$ & 3.083$\times 10^{-4}$ \\
$\nu=5, \; \omega = I_3$ & 2.614$\times 10^{-5}$ & 2.315$\times 10^{-5}$ & 2.177$\times 10^{-5}$ & 2.093$\times 10^{-5}$ & 2.183$\times 10^{-5}$ \\
$\nu=5, \; \omega = 2I_3$ & 8.125$\times 10^{-6}$ & 7.132$\times 10^{-6}$ & 6.729$\times 10^{-6}$ & 6.990$\times 10^{-6}$ & 7.327$\times 10^{-6}$ \\ \hline
\end{tabular}
\end{table}

The results are presented in Tables \ref{asym2} and \ref{asym3}. From these tables, it can be observed that the values of the empirical percentiles of the appropriately scaled test statistic tend to stabilize with the increase of $n_1$ and $n_2$. However, the parameters $\nu$ and $\omega$ greatly influence the values of the percentiles of the empirical distribution.
 }

\end{document}